\documentclass{aa}  
\usepackage{natbib}
\bibpunct{(}{)}{;}{a}{}{,} 
\usepackage{graphicx}
\usepackage{txfonts}
\usepackage{subfig}
\usepackage{soul}
\usepackage[dvipsnames]{xcolor}
\usepackage{amsmath}
\usepackage{hyperref}
\hypersetup{
    colorlinks=true,
    citecolor=blue,
    linkcolor=blue,
    filecolor=magenta,      
    urlcolor=cyan}
\usepackage{caption}
\usepackage{lipsum}
\usepackage{animate}
\usepackage[normalem]{ulem}

\newcommand{\luv}{L_{\rm UV}}
\newcommand{\lx}{L_{\rm X}}

\begin{document}

   \title{Quasars as Standard Candles}

   \subtitle{IV. Analysis of the X-ray and UV indicators of the disc-corona relation}

  \author{Matilde Signorini\inst{1,2,3}\thanks{\email{matilde.signorini@unifi.it}}, Guido Risaliti\inst{1,2}, Elisabeta Lusso\inst{1,2}, Emanuele Nardini\inst{2}, Giada Bargiacchi\inst{4,5}, Andrea Sacchi\inst{6}, \and Bartolomeo Trefoloni\inst{1,2}}
  
   \institute{Dipartimento di Fisica e Astronomia, Universit\`a di Firenze, via G. Sansone 1, 50019 Sesto Fiorentino, Firenze, Italy
         \and
             INAF - Osservatorio Astrofisico di Arcetri, Largo Enrico Fermi 5, I-50125 Firenze, Italy
         \and
            University of California-Los Angeles, Department of Physics and Astronomy, PAB, 430 Portola Plaza, Box 951547, Los Angeles, CA 90095-1547, USA
         \and
             Scuola Superiore Meridionale, Largo S. Marcellino 10, 80138 Napoli, Italy
         \and
             Istituto Nazionale di Fisica Nucleare (INFN), Sez. di Napoli, Complesso Univ. Monte S. Angelo, Via Cinthia 9, 80126, Napoli, Italy
         \and
			Center for Astrophysics | Harvard \& Smithsonian, 60 Garden Street, Cambridge, MA 02138, USA
             }

\titlerunning{Quasars as Standard Candles IV. X-ray and UV indicators.}
\authorrunning{M. Signorini et al.}

   \date{\today}

  \abstract
{A non-linear relation between quasar monochromatic luminosities at 2500~\AA\ and 2~keV holds at all observed redshifts and luminosities, and it has been used to derive quasar distances and to build a Hubble Diagram of quasars. The choice of the X-ray and UV indicators has so far been somewhat arbitrary, and has typically relied on photometric data.}
{We want to determine the X-ray and UV proxies that provide the smallest dispersion of the relation, in order to obtain more precise distance estimates, and to confirm the reliability of the X-ray to UV relation as a distance indicator.}
{We performed a complete UV spectroscopic analysis of a sample of $\sim$1800 quasars with SDSS optical spectra and {\it XMM-Newton} X-ray serendipitous observations. In the X-rays, we analysed the spectra of all the sample objects at redshift $z > 1.9$, while we relied on photometric measurements at lower redshifts. As done in previous studies, we analysed the relation in small redshift bins, using fluxes instead of luminosities.}
{We show that the monochromatic fluxes at 1~keV and 2500~\AA\ are, respectively, the best X-ray and UV continuum indicators among those that are typically available. We also find a tight relation between soft X-ray and Mg\,\textsc{ii}\,$\lambda2800$\AA\ line fluxes, and a marginal dependence of the X-ray to UV relation on the width of the Mg\,\textsc{ii} line.}
{Our analysis suggests that the physical quantities that are more tightly linked to one another are the soft X-ray flux at $\sim$1~keV and the ionizing UV flux blueward of the Lyman limit.  However, the ``usual'' monochromatic fluxes at 2~keV and 2500~\AA\ estimated from photometric data provide an almost as-tight X-ray to UV relation, and can be used to derive quasar distances. The Hubble diagram obtained using spectroscopic indicators is fully consistent with the one presented in previous papers, based on photometric data.}


\maketitle
%
\section{Introduction}
\label{intro}
Quasars are the most luminous persistent objects in our Universe. Their spectral energy distribution (SED) is complex; it goes from the radio to the X-rays, with the most intense emission emerging at optical--UV wavelengths \citep[e.g.][]{Sanders89,Richards06,Elvis2012}. The origin of this emission is attributed to accretion from an optically thick and geometrically thin disc around the central supermassive black hole \citep[SMBH,][]{SS73}.  
Since decades, the presence of a non-linear relation between the X-ray and UV luminosities of quasars has been observed \citep{Tananbaum79}. This relation is usually parameterised as $\log(\lx) = \gamma \log(\luv) + \beta$, where the slope is found to be $\gamma\simeq0.6$ over a wide range of redshifts and luminosities \citep[e.g.][]{Steffen06,Lusso10,Young2010}.
This relation must be based on the interaction between the accretion disc, which emits mainly in the UV, and the so-called X-ray corona, which consists of a hot-electron plasma. UV photons coming from the disc are up-scattered in the corona, where they reach X-ray energies. It is clear that such inverse-Compton mechanism can rapidly cool down the electron plasma, thus halting the production of X-rays. Since the X-ray emission of quasars is, instead, found to be persistent, there must exist a mechanism that refuels the corona with energy. Given that the \textit{engine} of a quasar is the infall of matter onto the central SMBH through the accretion disc, this is also where the energy that refuels the X-ray corona most likely comes from. The existence of the $\lx-\luv$ relation is therefore thought to be linked to such a mechanism, even if the exact physical process is still not completely understood. Modeling attempts have considered the reprocessing of the radiation from a non-thermal electron-positron pair cascade \citep{Svensson82}, buoyancy and reconnection of magnetic fields as a way to dissipate the gravitational power \citep{Haardt91,Haardt93,Svensson1994}, or the presence of a viscosity-heated corona, in which friction produces the heating \citep{Meyer2000}.

In addition to its relevance to quasars physics, the non-linearity of the X-ray to UV luminosity relation and the non-variability of its parameters make it possible to determine the luminosity distance of quasars, and therefore to use them as standard candles. This cosmological implication has been evident since the discovery of the $\lx-\luv$ relation. However, its application was hindered by the very high observed dispersion ($\sim$0.4 dex), which made the luminosity distance estimates too uncertain to be useful for cosmology. In recent years, it has been shown that most of this dispersion is due to observational issues, which can be largely removed by neglecting from the sample quasars that are affected by dust reddening, gas absorption, or Eddington bias \citep[for instance,][]{RL16_tight,Lusso20}. This way, the observed dispersion is significantly reduced to $\sim$0.20--0.25 dex. The resort to quasars as standard candles allows us to extend the Hubble diagram up to redshift values much higher than the ones that are achieved with supernovae Ia \citep{Scolnic18,Scolnic22}. This extended Hubble diagram is in agreement with the predictions of a flat $\Lambda$CDM cosmology up to redshift $z\sim$1.5. At the same time, it shows a 4$\sigma$ tension with the concordance model at higher redshifts \citep[e.g.][]{RL19_nature,bargiacchi21,bargiacchi22}.
Given the obvious relevance of these cosmological results, and the absence of a physical model explaining the X-ray to UV luminosity relation, it is fundamental to validate it observationally, ruling out any possible redshift dependence and systematic biases in the selection of the sample and/or in the flux measurements. A possible way to perform such checks has been proposed by other groups (e.g. \citealt{khadka2021}), based on the comparison of the data with several different cosmological models. 
However, it is impossible to perform a complete observational and cosmology-independent validation of a standard candle beyond the maximum redshift at which it has been validated by other distance indicators. Here, we use a different approach: in order to convince ourselves that the adoption of our method beyond $z=1.5$ (at lower redshift it has already been validated, by the comparison with SNIa \citep[see, e.g.,][]{Lusso20}) is well motivated, we want to further exploit our observational data set, in particular using the spectroscopic data, instead of just the photometric ones, for a better analysis of the X-ray to UV relation. 
In this paper, we will focus on the search of the optimal X-ray and UV indicators for the observed relation. In subsequent works we will focus on the analysis of the intrinsic dispersion of the relation (Signorini et al. 2023, in preparation), and on the analysis of the average spectral properties of ous sample as a function of redshift (Trefoloni et al. 2023, in preparation).

In previous studies, the 2500-\AA\ and 2-keV luminosities have been used as $\luv$ and $\lx$, respectively. The reasons behind this choice are mainly historical. Although the exact nature of the physical interaction between the disc and the corona is not completely understood, it seems unlikely to involve two monochromatic luminosities. Conversely, it is reasonable to believe that it should depend on the UV and X-ray emission over wider bands. Indeed, by adopting monochromatic luminosities for $\lx$ and $\luv$ to analyse the $\lx-\luv$ relation, we are simply choosing two \textit{proxies} of the overall emission. Therefore, we can ask ourselves which quantities work better to minimize the observed dispersion of the relation.
The main aim of this paper is to discuss the choice of the $\lx$ and $\luv$ indicators to establish whether it is possible to further reduce the observed dispersion, and so gain a better understanding of the physics behind the $\lx-\luv$ relation.

We also note that in most of our previous studies, the UV monochromatic luminosities have always been derived from photometric data. In principle, deriving luminosities from a spectroscopic analysis is a more precise method. This is true especially in the UV, as a complete spectral fitting allows us to accurately take into account all the emission lines, while in photometric estimates there might remain some contamination from line emission that is not separated from the continuum. It is then worth investigating the results of implementing UV measurements derived from a thorough spectroscopic analysis instead of the photometric ones.

The paper is structured as follows: in Section~2 we present the data sample and the products of our spectroscopic analysis; in Section 3 we make use of the spectroscopic sample to investigate the X-ray to UV relation using different proxies of the UV and X-ray fluxes; in Section~4 we discuss the Mg\,\textsc{ii} line width as a possible additional parameter of the relation. Section~5 presents the discussion of our results, consisting of two parts: (a) the analysis of the differences among the parameter values of the relation depending on the choice of the UV and/or X-ray proxies; (b) the implications for the use of quasars as cosmological distance estimators.
Our conclusions are summarized in Section~6.
Source luminosities are estimated by adopting a concordance flat $\Lambda$CDM cosmology with $H_0=70\, \rm{km \,s^{-1}\, Mpc^{-1}}$ and $\Omega_\mathrm{M}=0.3$.

\section{Data}
Our sample consists of 1764 quasars, all of which have available UV and X-ray data in public catalogues and are included in the sample published by \cite{Lusso20}. In particular, we considered in the said sample all the sources with available X-ray observations from the {\it XMM-Newton}satellite. More in detail, the sample is made of the following subsamples of the \cite{Lusso20} one: SDSS-4XMM (1644 objects), XMM-Newton $z \simeq 3$ (14 objects), and XXL (106 objects). Out of 1764, 772 objects have spectra obtained with the BOSS spectrograph, while the remaining have spectra acquired with the original SDSS one.
As discussed in the next Section, our analysis is performed by dividing the sample into small redshift bins. The high-redshift ($z\geq 4$) objects 
discussed in Sections 2.5 and 2.6 of \cite{Lusso20} are too sparse in redshift for the kind of analysis that we are interested in. This is why we excluded them from this analysis although they have \textit{XMM-Newton} observations available. \\ 

Detailed information regarding the sample can be found in \citet[][see their Section 2]{Lusso20}. Here we briefly recall that the selected objects are all radio-quiet and not flagged as broad absorption line (BAL) quasars. Furthermore, they have blue optical colours and steep X-ray spectra (photon index $\Gamma>1.7$), and they have deep enough X-ray observations to avoid biases towards brighter-than-average states. These properties make them suitable for a detailed analysis of the X-ray to UV relation, taking advantage of the homogeneity of both the source properties and the observational data.

\subsection{Optical--UV spectral analysis}

We performed the fitting procedure of each of the SDSS spectra using the software package \texttt{QSFit} \citep{Calderone17}. The S/N ratio per pixel for our sample is typically between 8 and 20 (pixels in SDSS/BOSS spectra are logarithmically spaced, with a width of  10$^{-4}$ dex).  For each object, we derived the continuum slope and the emission line properties (i.e. integrated line flux, rest-frame equivalent width, full-width at half-maximum, velocity offset). We also recorded the continuum monochromatic flux at different rest-frame wavelengths (1350 {\AA}, 2500 {\AA}, 3000 {\AA}, 4400 {\AA}, 5100 {\AA}). When one or more of these five rest-frame wavelengths did not fall in the range covered by the SDSS spectrum, the relative monochromatic fluxes were derived through the extrapolation of the best-fit continuum. A typical spectrum with its best-fit model is shown in Figure~\ref{exUV}, while a detailed description of the fitting procedure is provided in Appendix A. We note that the \texttt{QSFit} code returns results in luminosities, and not fluxes, assuming a standard $\Lambda$CDM cosmology with $H_0$= 70 km s$^{-1} $Mpc$^{-1}$, $\Omega_{\rm M}$ = 0.3 and $\Omega_{\Lambda}$ = 0.7. Given that we are interested in flux measurements, we derived them from the obtained luminosities assuming the same cosmology. All the objects also have photometric flux estimates, derived as described in \cite{Lusso20}. 

\begin{figure}
	\includegraphics[scale=0.3]{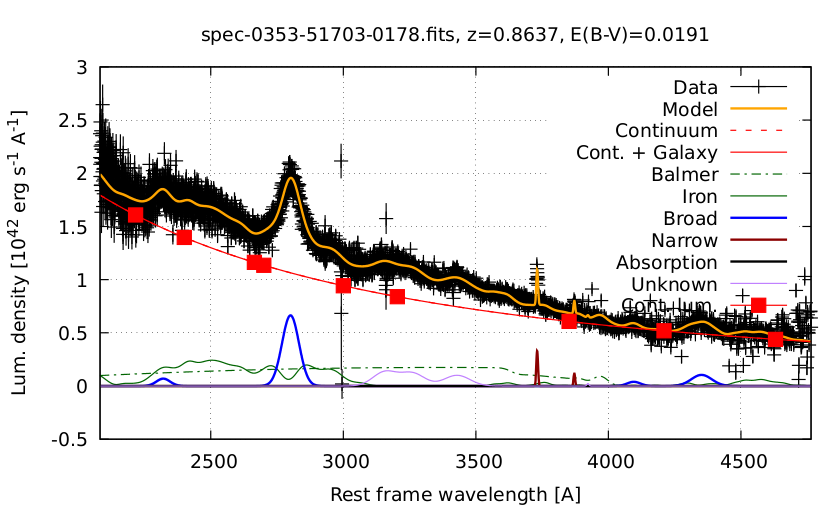}
	\centering
	\caption{Example of a UV spectrum together with the best fit (yellow line) and the various spectral components. The \texttt{QSFit} code assumes, for calculating luminosities, a standard $\Lambda$CDM cosmology with $H_0=70$ km s$^{-1}$ Mpc$^{-1}$, $\Omega_{\rm M}$ = 0.3 and $\Omega_{\Lambda}$ = 0.7 \citep{Calderone17}.}
	\label{exUV}
\end{figure}

\begin{figure}
	\includegraphics[scale=0.6]{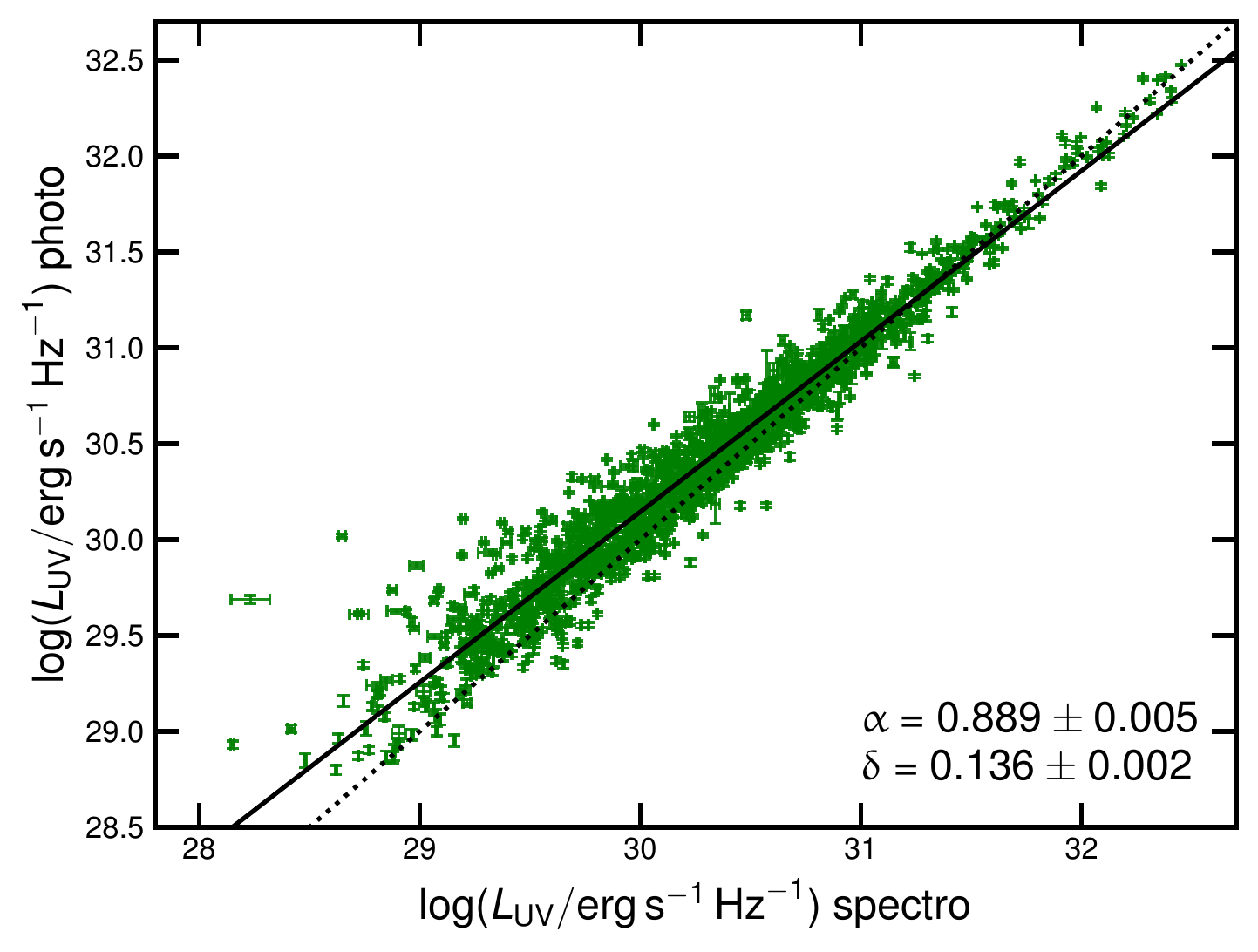}
	\centering
	\caption{Relation between the monochromatic luminosity at 2500~{\AA} derived from photometry as a function of the one obtained from the spectral analysis of our sample. The dotted line represents the one-to-one relation. The solid black line is the best-fit regression between these two parameters. The resulting slope and dispersion are reported in the figure.}
	\label{spph_comp}
\end{figure}

For completeness, Figure~\ref{spph_comp} presents the comparison between the monochromatic luminosity at 2500~{\AA} derived from photometry as a function of the one obtained from the spectral analysis of our sample. The dotted line represents the one-to-one relation, whilst the solid black line is the best-fit regression between these two parameters. The resulting slope and dispersion are also reported in the figure. Interestingly, the relation between the two monochromatic flux estimates is significantly non-linear. The discussion of this result is beyond the scope of this paper, and we will address it in a forthcoming work.

\subsection{X-ray spectral analysis}\label{xspecan}
Regarding the X-ray data, we have performed a complete spectral analysis and derived the photon index $\Gamma$ and the monochromatic fluxes for all the objects in our sample at a redshift higher than 1.9 (292 out of 1764). This is part of the ongoing effort to get better estimates of both the X-ray monochromatic fluxes that are used in the $\lx-\luv$ relation and the X-ray photon indexes $\Gamma$, which are key parameters in the sample selection to remove possibly obscured quasars.\footnote{The spectroscopically derived photon indexes $\Gamma$ are used to improve our sample selection, adopting the criterion of $\Gamma>1.7$. This introduces an inhomogeneity in the selection of objects at low and at high redshift. However, we found that the only differrence between the photometric and spectroscopic selection at $z>1.9$ is a slight decrease of the dispersion of the relation in the spectroscopically selected sample, without any systematic effects on the parameters of the relation.}
The spectroscopic fluxes for the $z>1.9$ are fully consistent with the photometric ones. The only improvement we obtained is a slight decrease of the dispersion of the UV to X-ray relation (see the discussion in \citealt{Sacchi22}). Therefore, we did not extend the (quite time consuming) spectroscopic analysis further. 
A detailed description of the X-ray spectroscopic analysis can be found in Appendix B. 

For the objects at redshift $z<1.9$, the rest-frame 2-keV flux is computed starting from the observer's frame fluxes at 0.5--2 keV ($F_{\rm S}$) and 2--12 keV ($F_{\rm H}$) tabulated in the 4XMM-DR9 serendipitous source catalogue. An analysis of simulated power-law spectra with typical {\it XMM-Newton} responses and effective areas shows that the monochromatic fluxes at 1 keV ($f_{1\,\rm keV}$) and at 3.45 keV ($f_{3.45\,\rm keV}$) are ``pivot points'' for the soft and hard band, respectively, i.e. the relation between these monochromatic fluxes and the total $F_{\rm S}$ and $F_{\rm H}$ fluxes is almost insensitive to the photon index $\Gamma$. Therefore, we estimated $f_{1\,\rm keV}$ and $f_{3.45\,\rm keV}$ from $F_{\rm S}$ and $F_{\rm H}$ assuming the same photon index used in the 4XMM catalogue ($\Gamma = 1.42$). We then used the power law defined by these two fluxes to compute both a new {\it photometric} photon index and the rest-frame monochromatic fluxes at different energies, as needed for our analysis (see the next sections). More details on this procedure are discussed in \cite{RL19_nature} and \cite{Lusso20}.

\section{Analysis of the X-ray to UV relation} 
The first step of our analysis of the X-ray to UV relation is a comparison of the results obtained adopting the new spectroscopic fluxes at the traditional reference wavelength of 2500 {\AA} with those based on our previous UV and X-ray flux values from photometric data. Before showing the results, we outline here the fitting method, which will be the same throughout the paper. 

As in our previous studies, we always performed our analysis by dividing the sample into a fixed number of redshift bins, such that for each bin, $\Delta \log z<0.1$. For this sample, we chose 11 bins in the 0.38--3.48 redshift range,  with a width of each bin of $\Delta \log z=0.08$. Doing so, the differences among luminosity distances for the objects in a given bin are negligible compared to the dispersion in the relation. In particular, we have checked with simulated data that the best-fit slope is always correctly recovered, provided that $\Delta \log z<0.1$. 
The relation can therefore be re-written in terms of fluxes as:
\begin{equation}
	\log(f_{\rm X})=\gamma \log(f_{\rm UV})+\beta',
		\label{law_f}
\end{equation}
where $\beta'$ is related to the normalization $\beta$ of the $\lx-\luv$ relation through the following expression:
\begin{equation}
	\beta'(z)=2(\gamma - 1) \log D_L(z)+(\gamma - 1) \log 4\pi+\beta.
		\label{betaz}
\end{equation}
With this choice, we can perform our analysis independently from the cosmological model used to compute the luminosities. This is fundamental as we want our results on the physical relation between UV and X-ray emission not to be biased as a consequence of the adopted cosmology, which is essential to subsequently implement quasars as cosmological probes. Furthermore, fitting the X-ray to UV relation in separate redshift bins allows us to investigate possible trends of its parameters with redshift, which is also crucial for cosmological applications. The parameters derived from the fitting procedure are $\gamma$, $\beta'$, and the intrinsic dispersion $\delta$. As in our previous works, we introduce the parameter $\delta$ because the observational errors on $\lx$ (or $f_{\rm X}$) and $\luv$ (or $f_{\rm UV}$) alone cannot explain the observed scatter of the $\lx-\luv$ (or $f_{\rm X}-f_{\rm UV}$) relation. For each bin, we also calculated the total dispersion $\delta_{T}$, which can be considered to be the square root of the quadratic sum of the total observational error on the fluxes and the \textit{intrinsic} dispersion $\delta$.

To perform the fit, a Bayesian approach of likelihood maximization was used: this allows us to estimate the parameters $\gamma$ and $\beta'$ while also taking the presence of the intrinsic dispersion $\delta$ into account, by modifying the likelihood function accordingly. Furthermore, using this approach we can obtain reliable estimates of the uncertainties on the parameters. We used the \textit{emcee} code, which is an implementation of Goodman $\&$ Weare’s Affine Invariant Markov chain Monte Carlo (MCMC) Ensemble sampler \citep{emcee13}. We performed a fit for each redshift bin, adopting a sigma-clipping at 3$\sigma$ in order to exclude the few strong outliers that might still be present after the quality-selection procedure. The sigma-clipping procedure removes $\sim$1\% of the sample, both when photometric and when spectroscopic data are used. The parameters of the relation ($\gamma$ and $\beta'$) do not change with the sigma-clipping; the only effect is on the dispersion parameter, which decreases by 0.02 dex for both samples. \\

Once established that no significant trend of the fitted parameters in the redshift bins is found, we derived the average value for the parameters $\gamma$, $\delta$, and $\delta_T$ by weighing each value for the number of objects in that bin. We remind that $\beta'$, instead, is expected to vary with redshift through its dependence on $D_L$ (see equation \ref{betaz}).

\begin{figure*}
    \centering
    \subfloat{{\includegraphics[width=8.75cm]{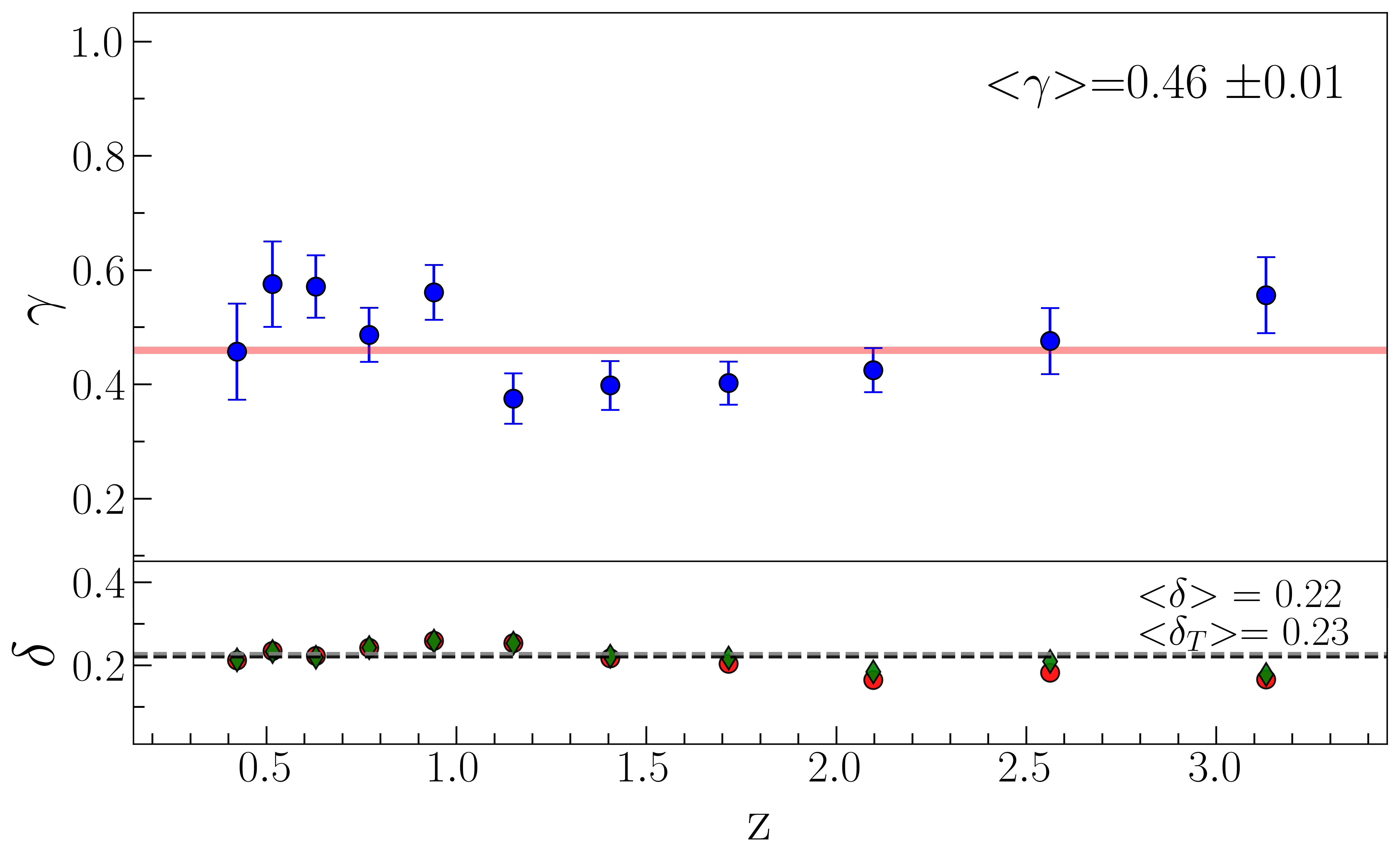} }}%
    \qquad
    \subfloat{{\includegraphics[width=8.75cm]{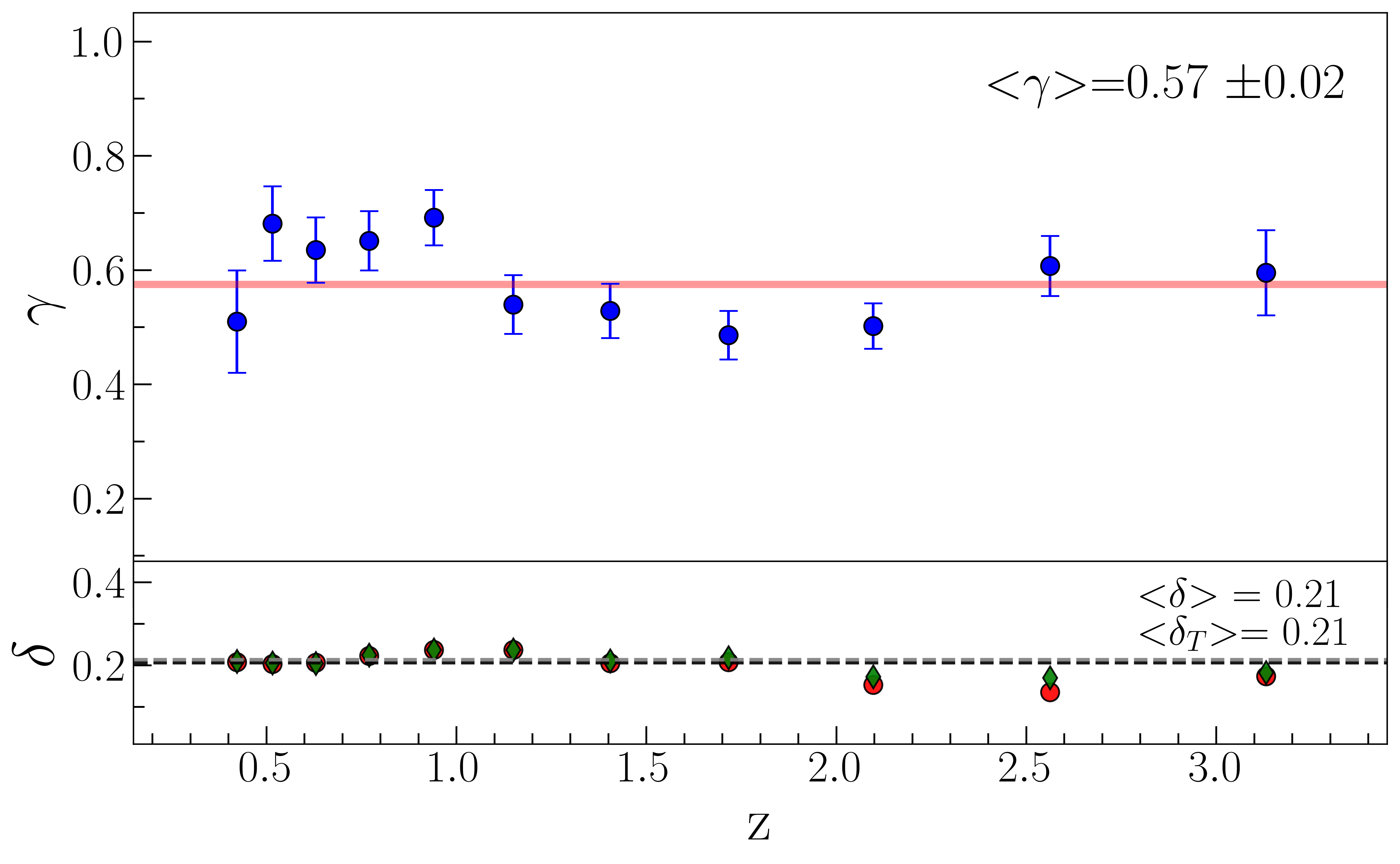} }}%
    \caption{Slope $\gamma$, total dispersion $\delta_T$ (as green diamonds), and intrinsic dispersion $\delta$ (as red circles) of the X-ray (2~keV) to UV (2500 \AA) flux relation in 11 narrow redshift bins. Left panel: results with the UV flux derived from spectroscopic data. Right panel: results with photometric UV fluxes.}%
    \label{fig:gammas}%
\end{figure*}

Following this procedure, we tested the X-ray to UV relation using the 2500 {\AA} monochromatic flux obtained in the spectroscopic analysis, while we keep assuming the photometric 2-keV monochromatic flux as $f_{\rm X}$. The slope parameter $\gamma$ of the flux-based relation is found to be always lower than in our previous analyses, where the photometric flux at 2500 {\AA} was used (\citealt{RL16_tight,LR17,RL19_nature,Lusso20}), with an average value of $\gamma=0.46$, as shown in Fig. \ref{fig:gammas}, left panel.
The slope does not show a systematic redshift trend. We tested this by fitting the 11 $\gamma$ values as a function of redshift with a line, $\gamma = m z + q$, finding the best fit values of the slope and intercept to be $m =  -0.02 \pm 0.03$ and $q = 0.49 \pm 0.05$. The slope $m$ is statistically consistent with zero, whilst the intercept $q$ is statistically consistent with the average value of $\langle \gamma\rangle=0.46$.

The redshift independence of the $\gamma$ parameter in the X-ray to UV  relation has already been thoroughly discussed in previous works, and it is now corroborated by employing spectroscopically-derived monochromatic fluxes.

Regarding the dispersion parameter $\delta$, the fact that we obtained a value of 0.22 dex, which is similar to (although slightly larger than) the one found when using photometric fluxes, is by itself an interesting result for our understanding of the physics of the X-ray to UV relation. If the ``true'' quantity behind the relation were the monochromatic luminosity at 2500 {\AA}, we would expect the intrinsic dispersion to be lower, as a spectroscopic fit is more accurately describing the true quasar continuum at a given wavelength when compared to the estimates we obtain from photometry. As this is not the case, it is possible that both spectroscopic and photometric fluxes are simply two different \textit{proxies} of the quasar UV emission, which are almost equivalently effective in terms of the tightness of the resulting relation. 

Starting from these considerations, we expanded our analysis in order to search for the optimal UV and X-ray indicators, under the assumption that both the UV and the X-ray spectra of quasars can be described by a power-law continuum over the range at issue. 
We expect the ``true'' physical relation to hold between the emission within a given frequency range in both the UV and the X-rays. For any frequency range, we can determine a {\it characteristic} frequency whose monochromatic flux would work as the best proxy according to the following argument: let us consider a power-law spectrum between the frequencies $\nu_1$ and $\nu_2$. Then, for a fixed total flux in the ($\nu_1,\nu_2$) interval, there will exist a frequency $\nu_{\rm C}$, which we call ``characteristic'', such that the monochromatic flux value $f_{\nu_{\rm C}}$ does not depend on the slope of the power law, as this is the point that divides the spectral range into two intervals with the same weight. In terms of frequency, we see that $\nu_{\rm C}\simeq\sqrt{\nu_1 \nu_2}$. This result is exact only if the spectral index is $-1$, i.e. $f_\nu \propto \nu^{-1}$, and the instrumental response in the ($\nu_1,\nu_2$) interval is flat. We argue that if we want to use a monochromatic flux as a proxy for the emission over a relatively wide wavelength range, the monochromatic flux at the ``characteristic energy'' would be the best choice. Indeed, it would be (nearly) independent of the specific value of the slope of the power-law spectrum and it would be directly related to the global emission in the ($\nu_1,\nu_2$) range. This argument works for the choice of both the UV and the X-ray proxies. 
 
To find the characteristic frequency, we considered that any generic frequency $\nu$ can be associated with the characteristic one through the expression:
\begin{equation}
	f_{\nu}=f_{\nu_{\rm C}}\left(\frac{\nu}{\nu_{\rm C}}\right)^{\Upsilon},
	\label{pp}
\end{equation}
where $\Upsilon$ is the slope of the power law, $\nu_C$ is the characteristic frequency and $f_{\nu_{\rm C}}$ is the corresponding flux. If we assume that equation \eqref{law_f} holds for the characteristic-energy flux $f_{\nu_{\rm C}}$ and we substitute equation \eqref{pp}, we see that if we use $f_{\nu}$, we also expect a correlation with the power-law slope $\Upsilon$. We expect to find a correlation with $\Upsilon$ when a non-characteristic flux is used, while we expect no correlation with $\Upsilon$ when the flux at the characteristic energy is used. Furthermore, the slope of such correlation would depend on the ratio between the frequency that we are using and the characteristic frequency. So we can define a modified version of the flux--flux relation that also takes into account the correlation with the slope $\Upsilon$:

\begin{equation}
	\begin{split}
		&\log(f_{\rm X})=\gamma \log(f_{\rm UV}) + \zeta \Upsilon+
		\beta', 
	\end{split}
	\label{piv_rel}
\end{equation}
where $\zeta$ is the slope of such correlation, which we can include in our fitting procedure as an additional parameter by modifying the likelihood function accordingly. We can use this relation to search for the characteristic energy in the UV (and likewise in the X-rays) of the $f_{\rm X}-f_{\rm UV}$ relation. The value of the parameter $\zeta$ is indeed related to the characteristic frequency, as:
\begin{equation}
\zeta = \gamma \log \left(\frac{\nu_{\rm C}}{\nu}\right).
\end{equation}

\subsection{UV monochromatic proxy}
Based on the aforementioned argument, we tested which monochromatic flux works as the best UV proxy by including the UV continuum slope as an additional parameter to the $\lx-\luv$ relation. As before, the fit was performed in 11 redshift bins and by using fluxes instead of luminosities. We used the same five different monochromatic fluxes derived from the spectroscopic analysis, as discussed in the previous section. We tested the relation \ref{piv_rel}, with the additional parameter $\zeta$ to be fitted and the UV slope as $\Upsilon$.

In terms of the parameter $\gamma$, we found average values in the range between 0.39 (for the 5100 {\AA} flux) to 0.45 (for the 2500 {\AA} and 3000 {\AA} fluxes). The significance of the additional parameter $\zeta$ is found to be very small for all the chosen UV proxies, although a tentative increase is seen at longer wavelengths.
The results are summarized in Table \ref{table_slope_spec}, and might have different explanations. It could be that a simple power law is not a sufficiently representative model for the optical--UV (1350--5100 \AA) continuum, so the ``characteristic energy argument'' does not completely hold, owing to, for example, some host-galaxy contribution at longer wavelengths. It could as well happen that the region of the disc that is truly physically linked to the X-ray corona is emitting at wavelengths shorter than (or across) the UV peak, so the optical--UV continuum slope is not the proper quantity to be taken into account.

\begin{table}[ht]
	\caption{Results of the search for the UV characteristic energy.} 
	\centering 
	\begin{tabular}{c c c c c} 
		\hline\hline 
		Wavelength & $\gamma$  & $\zeta$ & $\delta$ & $\delta_T$\\ [0.5ex] 
		\hline 
		1350 {\AA} & 0.43 $\pm$ 0.02  & 0.05 $\pm$ 0.02 & 0.23 & 0.23 \\ 
		2500 {\AA}  & 0.45 $\pm$ 0.02 & 0.05 $\pm$ 0.02 & 0.22 & 0.23\\
		3000 {\AA}  & 0.45 $\pm$ 0.02 & 0.05 $\pm$ 0.02 & 0.23 & 0.23\\		
		4400 {\AA} & 0.41 $\pm$ 0.02 & 0.09 $\pm$ 0.02 & 0.23  & 0.23\\
		5100 {\AA} & 0.39 $\pm$ 0.02 & 0.09 $\pm$ 0.02 & 0.23 & 0.24\\ [1ex] 
		\hline 
	\end{tabular}
	\label{table_slope_spec} 
\end{table}

Given these results, each of the five proposed optical--UV proxies could, in principle, work equivalently well.
The 2500 {\AA} wavelength falls within the SDSS spectra for the widest possible redshift range in our sample. This means that there are relatively few objects for which the monochromatic flux is derived through extrapolation, which might be less reliable that the directly measured ones. Indeed, the 2500 {\AA} flux provides the lowest dispersion, although marginally. We can thus consider this flux as the best proxy available.

We conclude this analysis with an additional fit required to compare the results obtained here to those published in previous papers on the X-ray to UV relation in quasars.
Throughout this paper, we have used for the first time  quantities derived from the spectroscopic analysis as $f_{\rm UV}$.  However, in all our previous works the UV monochromatic flux has been derived from the SDSS photometric data, through an interpolation procedure \citep{Lusso20}. We refer to this latter quantity as the ``photometric'' UV flux. 
In order to facilitate comparison, we fit the relation between fluxes with the 2-keV monochromatic flux as $f_{\rm X}$ and the photometric flux at 2500 {\AA} as $f_{\rm UV}$, within the same redshift range of 0.38--3.48 and over 11 redshift bins. We obtain a mean value of the slope $\gamma = 0.57 \pm 0.02$, and an intrinsic dispersion $\delta = 0.21$ dex. The results are shown in Fig. \ref{fig:gammas}, in the right panel.

\subsection{X-ray proxy}
The X-ray spectrum of quasars can also be described by a power law. Therefore, analogous arguments regarding the choice of the ``characteristic energy'' can be made, and we can investigate whether the 2-keV monochromatic luminosity is the best choice for $\lx$. The argument is the same as the one described in the previous section: if we are not using the characteristic flux as $f_{\rm X}$, we expect a dependence on the X-ray continuum slope, which in this case is described by the photon index $\Gamma$: 

\begin{equation}
	f_{2\,\rm keV}=f_{E_{\rm C}}\left(\frac{E_{2\,\rm keV}}{E_{\rm C}}\right)^{1-\Gamma}.
	\label{piv}
\end{equation}

We note that the X-ray spectral shape is much better described by a simple power law than the UV one, as there are no other prominent spectral features. 
Therefore, the determination of the ``characteristic energy" may be more straightforward than for the UV range.
The main difference with the previous section is that for the X-ray data, we still do not have complete spectral analysis for each object, so we rely on photometrically-derived estimates of both the monochromatic flux and the photon index for most of them. \\
We tested whether any dependence on the photon index $\Gamma$ is found when using the 2-keV monochromatic flux, implementing in our fitting procedure a modified version of equation \ref{piv_rel}:
\begin{equation}
\log (f_{\rm X}) = (\Gamma-1)\,\xi + \gamma \log(f_{\rm UV}) + \beta',
\end{equation}
where: 
\begin{equation}
\xi=\log\left(\frac{E_{\rm C}}{E}\right).
\label{xi}
\end{equation} 
We perform the analysis using the spectroscopically-derived flux at 2500 {\AA} as $f_{\rm UV}$. As before, we used 11 redshift bins, in the redshift range $z=0.38-3.48$. We obtain a mean value of the slope $\gamma = 0.47 \pm 0.02$, perfectly consistent with the previous values, and a significant $f_{\rm X}-\Gamma$ correlation parameter, with a mean value among the redshift bins of $\xi = -0.31 \pm 0.02$. The mean value of the intrinsic dispersion parameter gets lower, with $\delta = 0.19$ dex instead of 0.22 dex. 

The value of the parameter $\xi$ allows us to determine where the X-ray ``characteristic energy'' is located, according to equation \ref{xi}.
Given the obtained value $\xi = -0.31$, we infer that the characteristic energy should be located at $\sim$1 keV. 
The characteristic energy depends on the physical extent of the X-ray power law, which is not well known. Considering the minimum and the maximum energy of the (unknown) optimal X-ray range for the relation, so that the characteristic energy $E_{\rm C} \sim \sqrt{E_{\rm min} E_{\rm max}}$, our estimate of $E_{\rm C}\sim1$ keV might mean, for example, $E_{\rm min}\sim0.01$ keV and $E_{\rm max} \sim100$ keV, or $E_{\rm min}\sim0.1$ keV and $E_{\rm max}\sim10$ keV.  
Employing both the 2-keV flux and the parameter $\Gamma$ in the relation mimics the employment of the 1-keV flux, which is almost insensitive to the exact value of $\Gamma$. 

In order to test our assumption that the 1-keV energy is the X-ray ``characteristic energy'', we derived, from photometric data, the monochromatic flux at 1 keV for all the sources in our sample; for objects at redshift higher than 1.9, we derived it from the fit of the X-ray spectrum. We then performed again the aforementioned analysis. We found no significant correlation with the photon index $\Gamma$, with values of $\xi$ consistent with zero. This confirms that we have found the true ``characteristic energy''. Furthermore, by adopting the combination of the 1-keV flux as $f_{\rm X}$ and the 2500 {\AA} spectroscopic flux as $f_{\rm UV}$, we obtain a mean slope value of $\gamma=0.46\pm0.01$ (consistent with the previous estimate), a value of $\xi$ consistent with zero, and consistent values of the intrinsic and total dispersion parameters, with $\delta=0.18$ dex and $\delta_T=0.21$ dex.

We recall that allowing for an intrinsic dispersion parameter $\delta$ is necessary because the scatter of the observational points in the $f_{\rm X}-f_{\rm UV}$ relation cannot be entirely justified by the errors on the observed fluxes. Given that the rest-frame 1-keV emission is not between the two X-ray pivot points (see Section \ref{xspecan}), and that with increasing redshift it gets progressively farther away from the observed energy range, the 1-keV fluxes are less precisely measured, with a mean observational error higher than the one obtained for the 2-keV monochromatic fluxes estimates. Consequently, it is possible that part of the observed decrease of the intrinsic dispersion with the 1-keV flux as $f_{\rm X}$ is due to an increase of the mean observational error on the 1-keV fluxes. However, when adopting the 1-keV flux instead of the 2-keV one, we also notice a reduction of the total dispersion $\delta_T$. We thus conclude that, although the 1-keV flux is a less precise proxy in terms of its associated observational error, it can still be stated that, physically, it is a more accurate one.

To sum up, in this section we investigated the choice of the UV and X-ray proxies of the $\lx-\luv$ relation. We argued that the best choice consists, in both cases, in the monochromatic flux whose frequency acts as the ``characteristic energy'' of the spectral emission. Regarding the UV side of the relation, we found no clear trend as a function of the continuum slope, and therefore it can be stated that the monochromatic flux at 2500 {\AA} is a good choice, as it provides the smallest intrinsic dispersion and it can be directly observed in optical--UV spectra for a wide redshift range.
Regarding the X-ray side of the relation, we found a significant correlation with the photon index $\Gamma$ when the 2-keV flux is used as $f_{\rm X}$. The observed dispersion decreases when the latter dependence is taken into account. From this, we can deduce that the X-ray characteristic energy should fall around 1 keV. The adoption of the 1-keV flux, indeed, results in no correlation with the photon index $\Gamma$, and a lower dispersion too. However, the photometric estimates of the 1-keV fluxes are significantly less precise than the ones at 2 keV for our sample. 

In Figure \ref{gamma_2,500spec_1kev} we show the results in terms of the parameter $\gamma$ as a function of redshift, with the (spectroscopic) 2500 {\AA} monochromatic flux as $f_{\rm UV}$ and the monochromatic flux at 1 keV as $f_{\rm X}$. In Figure \ref{lenzuolata} we also show the $f_{\rm X}-f_{\rm UV}$ relation in the 11 redshift bins. With these choices, both the intrinsic and the total dispersion of the relation between fluxes are reduced. 
We checked again for the redshift dependence of the slope parameter $\gamma$, again by fitting the 11 values with a line as a function of redshift. We found the best-fit slope to be $m = -0.04\pm0.03$. Such value is compatible with zero at 1.3$\sigma$ and, overall, very small.
We stress again that the non-dependence of the $\gamma$ parameter with redshift is fundamental to show that (i) the physical mechanism behind the relation is the same at all the observed redshifts and (ii) we can safely use quasars as standard(izable) candles for cosmology. 

\begin{figure}
	\includegraphics[scale=0.32]{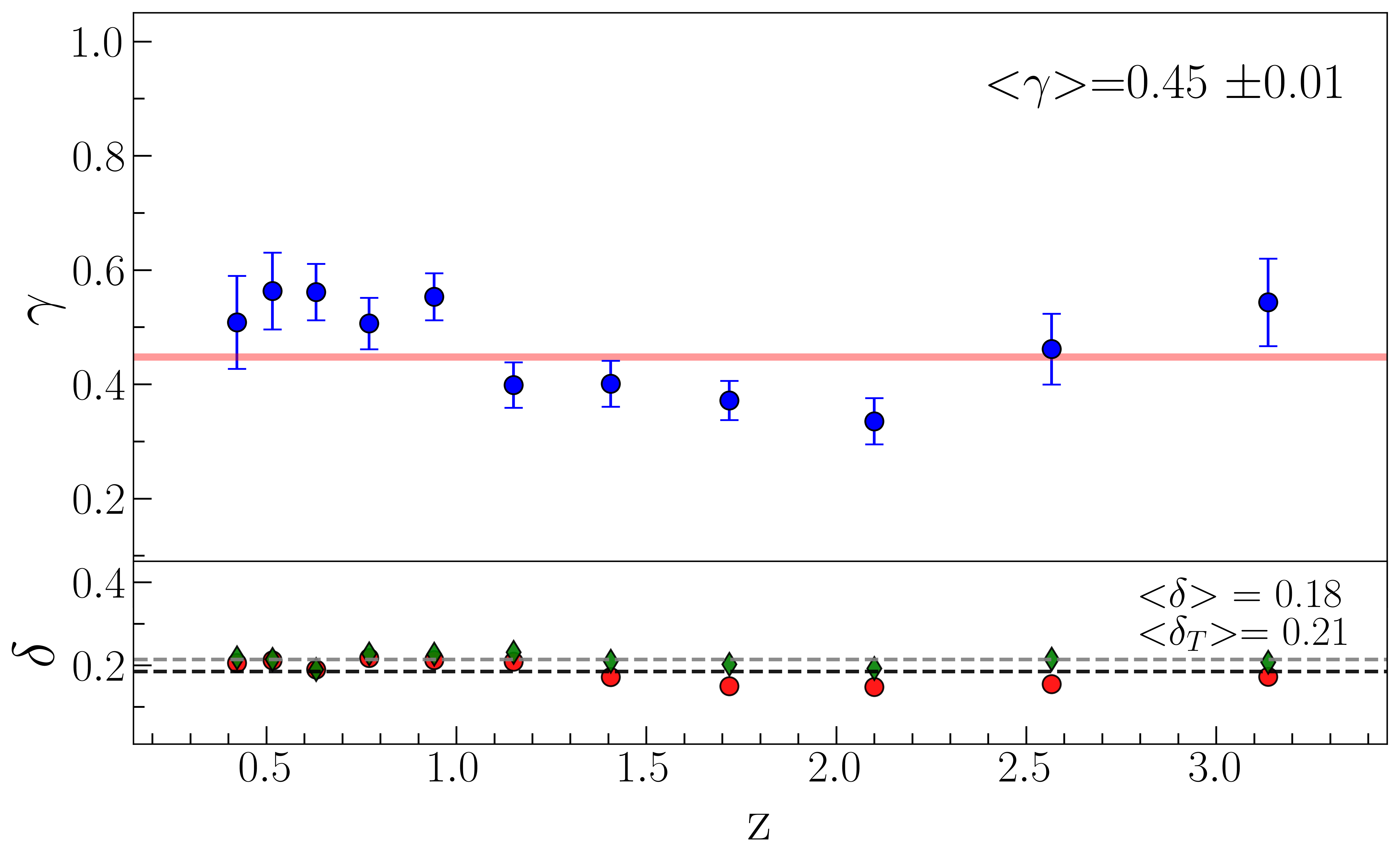}
	\centering
	\caption{Slope parameter $\gamma$ as a function of redshift, as obtained by fitting equation \ref{law_f} when the monochromatic flux at 2500 {\AA} derived from the spectroscopic analysis is used as $f_{\rm UV}$ and the monochromatic flux at 1 keV is used as $f_{\rm X}$. The mean value of $\gamma$ is 0.45$\pm$0.01, and we can see that there is no clear trend with redshift.}
	\label{gamma_2,500spec_1kev}
\end{figure}

\begin{figure*}
	\includegraphics[scale=0.73]{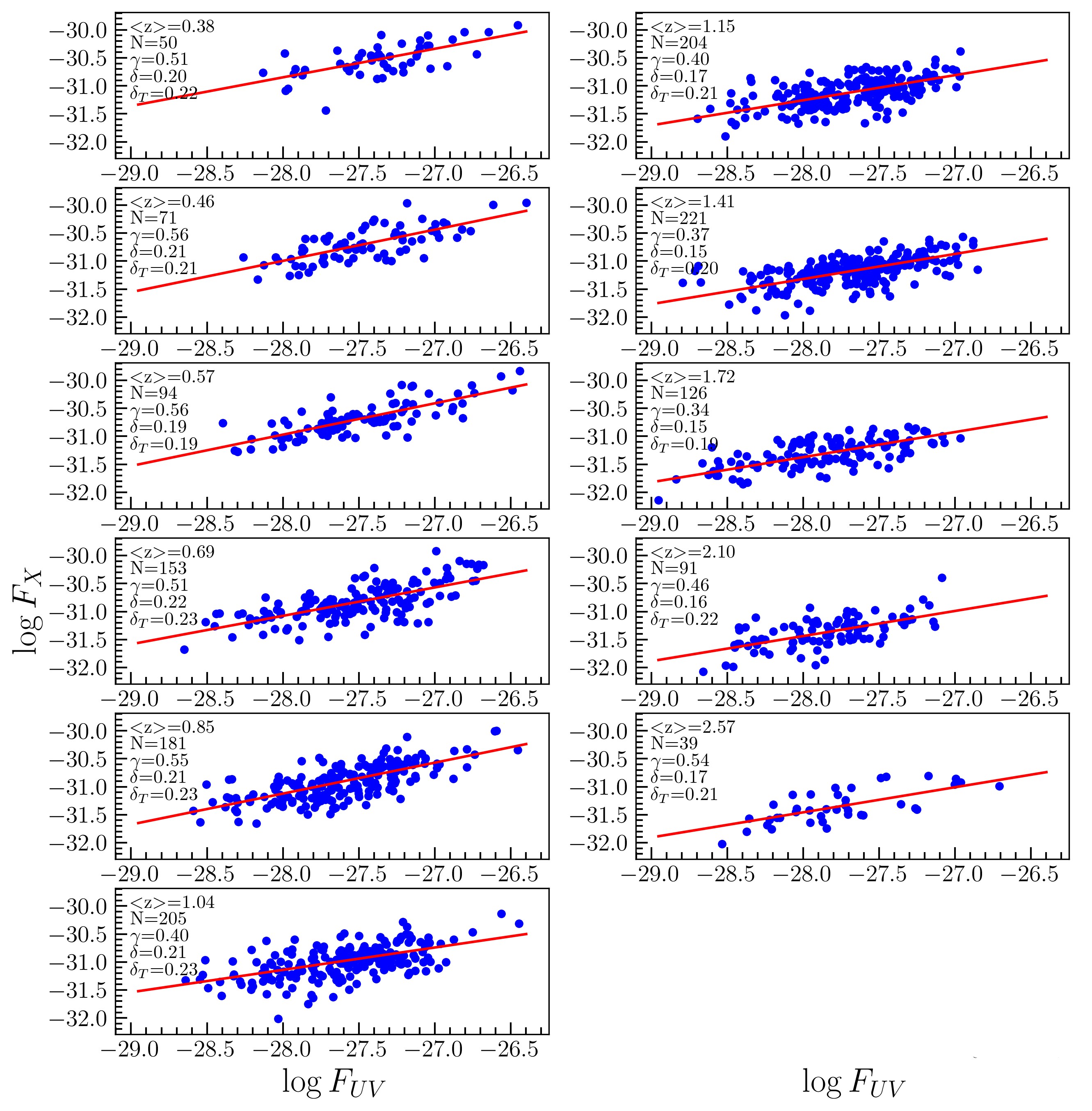}
	\centering
	\caption{$f_{\rm X}-f_{\rm UV}$ relation in 11 redshift bins, obtained by using the 2500 {\AA} monochromatic flux from the spectroscopic analysis as $f_{\rm UV}$ and the monochromatic flux at 1 keV as $f_{\rm X}$. The best-fit parameters (slope and dispersion), the total dispersion, the number of objects, and the average redshift are also reported.}
	\label{lenzuolata}
\end{figure*}

\label{section_proxy}

\subsection{Mg\,\textsc{ii} line flux as UV proxy}
The fluxes of quasar recombination lines depend on the intensity of the photoionizing continuum. As such, these fluxes could constitute  good indicators of the disc primary emission. We chose to test this hypothesis by analyzing the $f_{\rm X}-f_{\rm UV}$ relation using the 1-keV monochromatic flux as $f_{\rm X}$ and the flux of the Mg\,\textsc{ii}\,$\lambda$2800{\AA} emission line as $f_{\rm UV}$.

The Mg\,\textsc{ii}\,$\lambda$2800{\AA} resonant line \citep{Netzer80, Krolik88} requires the second ionization of magnesium by photons with energies exceeding 15.035 eV, i.e., with a wavelength shorter than 824.6 {\AA}. Therefore, it is sensitive to the UV continuum blueward of the Lyman limit, which is not accessible through direct observations. Furthermore, its high equivalent width allows relatively precise flux measurements, and its rest frame wavelength makes it observable in most of the redshift range of our spectroscopic sample. This combination of properties makes it the best candidate for our analysis, as it allows us to investigate whether the $\sim$800 {\AA} band is a
more appropriate indicator of the disc emission than the 1350--5100 {\AA} continuum. Alternative possibilities are the C\,\textsc{iv}\,$\lambda$1549{\AA} line at higher redshifts \citep{lussoCIV}, or the H$\beta$\,$\lambda$4861{\AA} line at lower redshift. The latter is indeed observable only up to $z\sim0.8$. 

\begin{figure}
	\includegraphics[scale=0.32]{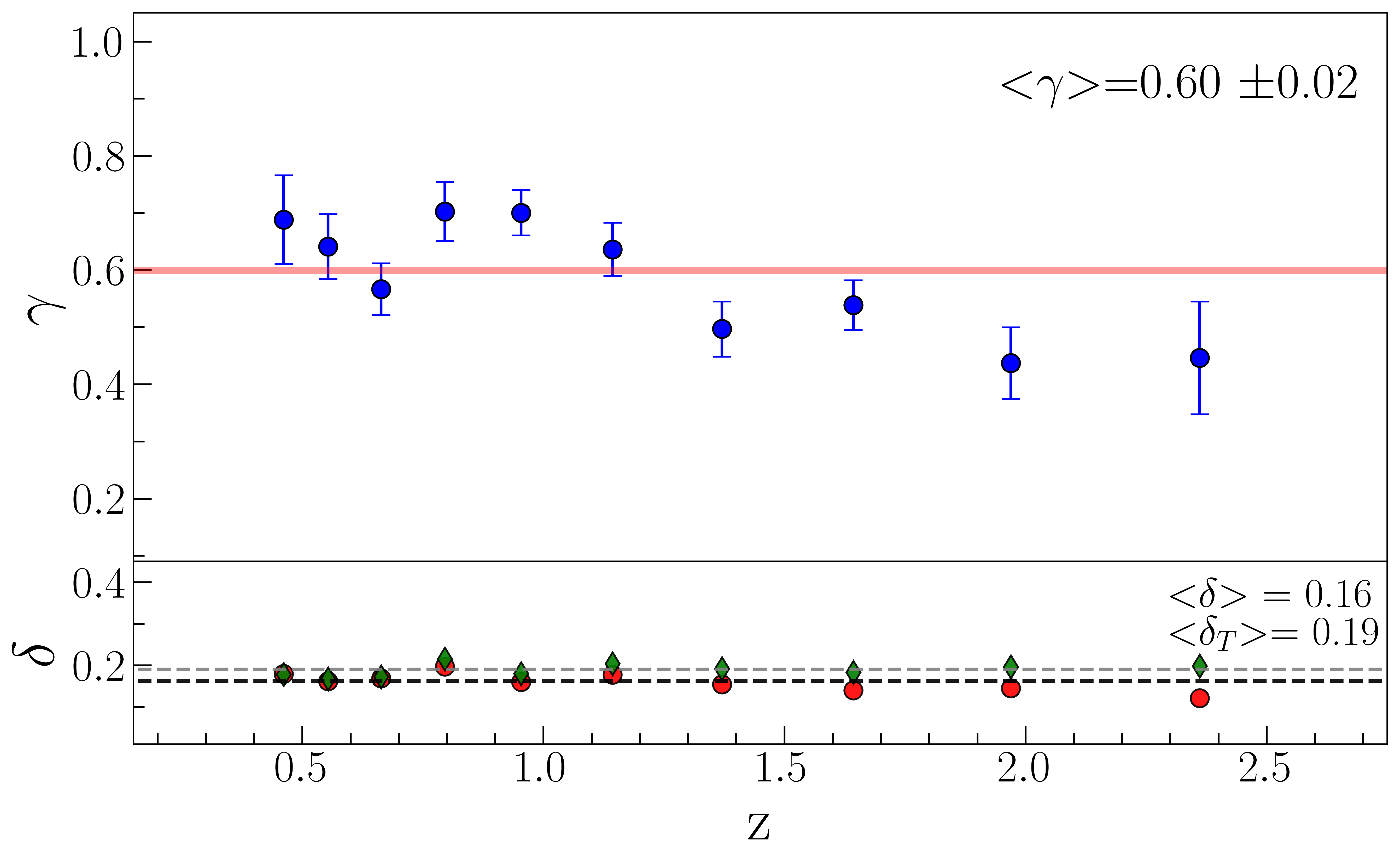}
	\centering
	\caption{Parameter $\gamma$ as a function of redshift, resulting from the fit of equation \ref{law_f} when the Mg\,\textsc{ii} emission line flux is used as $f_{\rm UV}$ and the monochromatic flux at 1 keV is used as $f_{\rm X}$. The mean value of $\gamma$ is 0.60 $\pm$ 0.02, and we can see that there is no clear trend with redshift.}
	\label{gamma_mg_1kev}
\end{figure}

The objects with spectra obtained with the SDSS spectrograph have Mg\,\textsc{ii} emission up to redshift 2.28, while the ones with spectra obtained with the BOSS spectrograph have  Mg\,\textsc{ii} emission up to redshift 2.71. However, the fitting procedure of the BOSS objects in the redshift range $2.58 < z < 2.71$ returned a bad quality flag for the Mg\,\textsc{ii} line, suggesting that the line parameters are not reliably constrained when this is located at the extremity of the spectrum. We, therefore, restricted the analysis to the redshift range $0.38<z<2.58$, using 10 redshift bins. We obtain a mean value of the slope $\gamma=0.60 \pm 0.02$, a mean dispersion parameter $\delta = 0.16$ dex, and a mean total dispersion $\delta_T = 0.19$ dex.
We notice that the Mg\,\textsc{ii} line flux as a UV flux indicator gives a higher value of the slope parameter when compared with the spectroscopically-derived monochromatic flux at any tested optical--UV wavelength, yet the dispersion is significantly lower. In Figure \ref{gamma_mg_1kev} we show the results of the fitting procedure using the Mg\,\textsc{ii} line flux as $f_{\rm UV}$ for the 10 redshift bins. We checked the dependence of the $\gamma$ parameter as a function of redshift with linear regression. The fit returns a $\gamma-z$ relation with a slope of $m = -0.11\pm0.04$. Contrary to the previous two cases, there is evidence for a redshift trend, even if not a very strong one. We notice however that the analysis with the Mg\,\textsc{ii} line is limited to a smaller maximum redshift, and the results in the common range are similar to those in Figs.~3 and 4 for the relations with the UV continuum. 
Another issue that might be affecting the Mg\,\textsc{ii} results are differences between objects with data coming from the SDSS spectrograph and the BOSS spectrograph. While we find no differences in the $\gamma$ values and redshift dependence when comparing BOSS and SDSS data for the continuum, differences are found when comparing Mg\,\textsc{ii} data. When fitting the $f_{\rm X}-f_{\rm UV}$ relation with only SDSS objects, we find an average slope of $\gamma=0.64\pm0.03$ and an even more significant redshift trend, so that, when fitting the slope $\gamma$ as a function of redshift with a line, we find a slope coefficient of $m = -0.19\pm0.05$. When using only the BOSS data, instead, we find $\gamma=0.61\pm0.03$ and no significant redshift trend, with $m =0.08\pm0.06$.

\section{Mg\,\textsc{ii} line width as a possible additional parameter}
The widths of emission lines are independent observables with respect to continuum or line fluxes, and carry complementary information, such as the gas rotational velocity or its outflow velocity. These properties are in turn related to the physical parameters of the primary source (black hole mass, accretion rate, etc.). For this reason it is interesting to investigate whether the X-ray to UV relation is also dependent on line widths, as already extensively discussed in \cite{LR17}. To perform this test, we employed in our fit the modified relation:
\begin{equation}
	\begin{split}
		\log(f_{\rm X})=&\gamma \log(f_{\rm UV}) + \eta \log(\rm FWHM_{Mg\,\textsc{ii}})+ \beta'',
	\end{split}
	\label{leggefw}
\end{equation}
where $f_{\rm X}$ is the monochromatic X-ray flux at 1 keV, and FWHM$_{\rm Mg\,\textsc{ii}}$ is the FWHM of the Mg\,\textsc{ii} emission line that we inferred with the UV spectral fitting procedure. 
The goals of this analysis are (a) to use possible additional correlations to decrease the intrinsic dispersion of the relation, and (b) to understand whether these correlations have an effect on the use of the X-ray to UV relation to estimate quasar distances.
The analysis is carried out in the 0.38--2.58 redshift range, where the Mg\,\textsc{ii} line is available, with 10 bins. 

When using the spectroscopically-derived monochromatic flux at 2500 {\AA} as $f_{\rm UV}$, we found a correlation which is highly significant, with a mean value of the parameter $\eta$ equal to $\eta = 0.25 \pm 0.04$. The observed intrinsic dispersion is  $\delta = 0.18$ dex and the total dispersion is $\delta_T=0.20$ dex. 
We note that the $\delta$ and $\delta_T$ parameters that we obtain are comparable with the values presented in the previous Section where the dependence on the FWHM was not taken into account.
It might be that the share of dispersion due to neglecting the FWHM correlation is too little for this effect to emerge, and/or that the dynamical range of the FWHM values is too small for the additional correlation to impact significantly on the global observed dispersion.
Regarding the parameter $\gamma$, we obtain a result consistent with the previous values, i.e.~$\gamma=0.45\pm0.01$. 

Using $f_{\rm Mg\,\textsc{ii}}$ as $f_{\rm UV}$, instead, we found no correlation with the FWHM: we obtain $\eta=-0.03 \pm 0.03$, consistent with zero. This suggests that when $f_{\rm Mg\,\textsc{ii}}$ is employed most of the significance of the correlation between $f_{\rm X}$ and line properties is already embedded in the $f_{\rm X}-f_{\rm Mg\,\textsc{ii}}$ relation, so that 
considering also a dependence on the FWHM does not add any new information. 

Given the high significance of the dependence of the X-ray flux on the FWHM parameter, the obvious next step of the analysis is to check for a possible correlation between the UV flux indicator and the FWHM. This correlation could in principle be relevant to the inversion of the X-ray to UV relation to obtain quasar distances and, in the worst case, introduce some redshift-dependent bias. In order to perform this check we need to compute the covariance matrix for the fit of Equation \ref{leggefw}. However, given the small dynamical range of the quantity log(FWHM) and the relatively small number of objects in each single redshift bin, we cannot obtain significant estimates of the correlation between UV flux and FWHM in the redshift-resolved analysis. For this reason, we carried out the fitting procedure using luminosities instead of fluxes (derived with a standard flat $\Lambda$CDM cosmology) and considering the whole sample together, using the monochromatic 1-keV luminosity as $\lx$, the spectroscopically-derived 2500 {\AA} monochromatic flux as $\luv$, and the FWHM of the Mg\,\textsc{ii} line. We then derived and normalized with respect to the first term the covariance matrix of $\gamma$ and $\eta$, and we diagonalized it, with the goal of finding the linear combination of $\luv$ and FWHM$_{\rm Mg\,\textsc{ii}}$ needed in order to have variables that are independent of one another in our fitting procedure. We found the matrix of the eigenvectors to be:
\begin{equation*} 
	\begin{pmatrix}
		-0.999 & -0.045  \\
		0.045 & -0.999
	\end{pmatrix},
\end{equation*}
which, multiplied by the variable column vector ($\luv$, FWHM$_{\rm Mg\,\textsc{ii}}$), gives us as new variables:
\begin{equation}
	\begin{split}
		& X = -0.999\,\luv -0.045\,{\rm FWHM_{Mg\,\textsc{ii}}}\\
		& Y = 0.045\,\luv -0.999\,{\rm FWHM_{Mg\,\textsc{ii}}}.
		\label{blabla}
	\end{split}
\end{equation}

These new variables are independent from one another: if we repeat the described procedure using $X,Y$ instead of $\luv$, FWHM$_{\rm Mg\,\textsc{ii}}$ we find a diagonal covariance matrix. 

This result also implies that we can implement a modified X-ray to UV flux--flux relation with the addition of the FWHM when this is available, and keep using the standard relation otherwise.
We performed this additional test in the whole redshift range of our sample, 0.38--3.48, again with 11 redshift bins, by modifying the likelihood accordingly. Using the photometric 2500 {\AA} flux as $f_{\rm UV}$ and the 1 keV flux as $f_{\rm X}$, we obtain $\gamma=0.56\pm0.01$, $\delta=0.16$ dex, $\delta_T=0.19$ dex, $\eta=0.25\pm0.04$. Therefore, with this mixed likelihood, we get a slight decrease of the intrinsic and the total dispersion.\\
In \cite{LR17}, the authors proposed a \textit{toy model} for the $\lx-\luv$ relation, where the relation 
arises from the dependence of both quantities on the black hole mass $M_{\rm BH}$ and the black hole accretion rate $\dot{M}_{\rm BH}$. The said model predicts the slope between the X-ray luminosity (or flux) and the FWHM of the Mg\,\textsc{ii} line to be $\hat{\eta} = 0.44^{+0.21}_{-0.19}$, 
and they found, for a sample of 545 quasars, an observed slope between the X-ray luminosity and the Mg\,\textsc{ii} FWHM equal to $\eta_{\rm LR17} = 0.54\pm0.07$, consistent with the toy model. Our result, $\eta = 0.25\pm0.04$, is also consistent with the prediction of the toy model. It significantly differs from the \cite{LR17} value, but we note that they performed the analysis (i) for the sample as a whole and not in redshift bins, using $\Lambda$CDM derived luminosities, and (ii) using the photometric 2500 {\AA} luminosity as $L_{\rm UV}$ instead of the spectroscopic one. If we perform the analysis in the same way, using the photometric luminosities for our sample, we obtain  $\eta_{\rm phot} =0.45\pm0.04$, statistically consistent with the \cite{LR17} result. \\
However, the consistency between our results and the toy-model presented in \cite{LR17} is not whole. The model, indeed, implied also the presence of a $L_{\rm UV}-$FWHM relation, while we find the two quantities to be uncorrelated. At the same time, it might be that the small dynamic range of the Mg\,\textsc{ii} FWHM does not allow us to observe the correlation.

Since the observational dependence of the X-ray flux on the Mg\,\textsc{ii} FWHM line may be due to a {\em physical} dependence on the black hole mass, we performed an additional check by repeating our whole analysis for two subsamples with black hole mass estimates $M_{BH}<10^{8.9} M_\odot$ and $M_{BH}<10^{8.9} M_\odot$, respectively. We did not find any significant differences between these two subsamples. This suggests that the parameters of the relation do not directly depend on the black hole mass.

We finally have to consider that all the results that we have discussed in this subsection might be dependent on the cosmological model used to compute luminosities from fluxes. To test this possible dependence, we followed a cosmographic approach to fit the quasar Hubble diagram, as described in \cite{bargiacchi21}.
We found analogous results in terms of covariance matrices, eigenvectors and $\luv-{\rm FWHM}$ dependence. 

\section{Discussion}
The main results obtained through the various fits presented in the previous sections are: \\
1) The slope $\gamma$ and the intrinsic dispersion $\delta$ of the X-ray to UV relation obtained by using spectroscopically-derived 2500 {\AA} fluxes are, respectively, $\gamma\sim0.46$ and $\delta\sim0.22$ dex.\\
2) The monochromatic flux at 2500 {\AA} can be considered as the best available proxy of the disc emission, given its observational coverage over a broad redshift range.\\
3) The best monochromatic indicator of the X-ray emission in the X-ray to UV relation (i.e. the one providing the smallest intrinsic dispersion) is the 1-keV flux.\\
4) The intrinsic dispersion obtained adopting the Mg\,\textsc{ii} line flux as UV proxy is smaller than that obtained with monochromatic continuum fluxes ($\delta_{\rm Mg\,\textsc{ii}}\sim0.17$ dex). The slope of the relation is instead significantly steeper ($\gamma_{\rm Mg\,\textsc{ii}}\sim0.60$) than in the spectroscopic case.\\
5) When the ``photometric estimate'' of the UV monochromatic flux is adopted, the best fit parameters (both slope and dispersion) are consistent with those found using the Mg\,\textsc{ii} line flux as UV proxy.\\
6) When the logarithm of the FWHM of the Mg\,\textsc{ii} line is added to the relation, a statistically significant correlation is found, with a negligible decrease of the total intrinsic dispersion. The Mg\,\textsc{ii} FWHM and UV flux parameters are not statistically correlated. 

Here we discuss the interpretation and the main consequences of these results.
The most remarkable result of our analysis is arguably that the spectroscopically-derived 2500 {\AA} monochromatic flux delivered a significantly lower $\gamma$ value when used as $f_{\rm UV}$ compared to the Mg\,\textsc{ii} line flux.
This trend can be explained as a consequence of the non-linear relation between the emission-line equivalent width (EW) and the luminosity of the quasar continuum, known as the ``Baldwin effect'' \citep{Baldwin77}. 
We analysed the relation between the Mg\,\textsc{ii} luminosity and the monochromatic luminosity at 2500~{\AA} derived from the spectroscopic analysis of our sample and we obtained a slope of $\sim$0.8. 
In order to fit the whole sample simultaneously, for this analysis we used luminosities instead of fluxes, assuming a standard flat $\Lambda$CDM model with $\Omega_{\rm M}$=0.3 and $H_0 $ = 70 km s$^{-1}$ Mpc$^{-1}$. We checked that the results are not significantly dependent on the choice of the cosmological model. 

If we now consider the relation between the X-ray and UV luminosities as $\log(L_{2\,\rm keV}) = \gamma \log(L_{2500}) + \beta$, with $\gamma = 0.46$ as shown in Section 2, and we use the $\log(L_{2500})-\log(L_{\rm Mg\,\textsc{ii}})$ mentioned above,  we obtain a slope of the  $\log(L_{2\,\rm keV})-\log(L_{\rm Mg\,\textsc{ii}})$ of $0.46/0.8 = 0.58$, which is fully consistent with the slope of the relation that we obtain when using the Mg\,\textsc{ii} luminosity (or flux) as $L_{\rm UV}$, as shown in Section 4. Therefore, we conclude that the reason behind different X-ray to UV relations when shifting from continuum UV proxies to line proxies 
is associated with the presence of the Baldwin effect itself.  

Another result related to the adoption of the Mg\,\textsc{ii} as UV proxy is the smaller intrinsic dispersion $\delta$ of the X-ray to UV relation with respect to the monochromatic continuum indicators. Considering that this line requires an ionizing continuum at wavelengths shorter than 824~{\AA}, a possible interpretation of this result is that an even tighter relation must exist between the X-ray flux and the UV flux blueward of the Ly limit. In this scenario, the monochromatic fluxes at optical or near-UV wavelengths are all ``secondary indicators'' with a similar relation to the primary one.

The slope parameter that we obtain when using the ``photometric'' UV flux has a less obvious interpretation. The photometric flux is a complex UV proxy, as it contains contributions from both the quasar continuum and line emissions. Moreover, even if it is formally a monochromatic quantity, it is derived from the wide-band photometric fluxes. We notice that its value is similar to the one obtained using the Mg\,\textsc{ii} line as UV proxy, and that the dispersion of the relation is marginally better than the one obtained with truly monochromatic fluxes. This suggests that the combined information used to derive the photometric flux is similar to that contained in the Mg\,\textsc{ii} flux, and is more closely related to the UV emission at $\lambda\sim800$~{\AA} than the monochromatic fluxes.

\begin{figure}
	\includegraphics[scale=0.6]{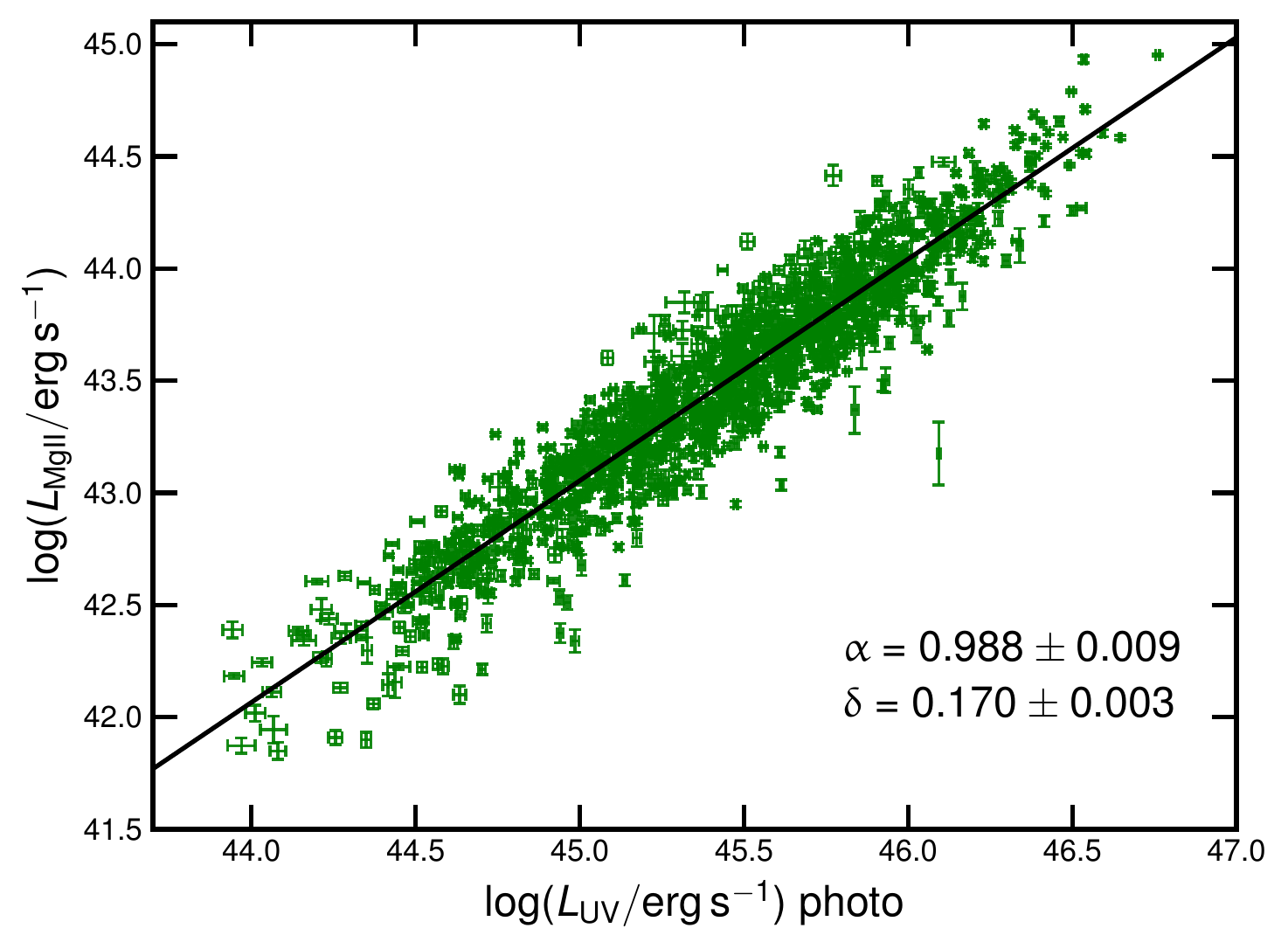}
	\centering
	\caption{Relation between the luminosity of the Mg\,\textsc{ii} line and the  luminosity at 2500~{\AA} derived from photometry.}
	\label{phmgii_comp}
\end{figure}

In Fig. \ref{phmgii_comp} we show the relation between the luminosity of the Mg\,\textsc{ii} line and the photometric luminosity at 2500~{\AA}. The slope is statistically consistent with unity, which explains the similar behaviour of these two quantities when used as $f_{\rm UV}$ proxies in the X-ray to UV relation.

\section{Cosmological application}
The most relevant result of our analysis concerning the use of quasars as standard candles through the X-ray to UV relation is the confirmation of such relation with spectroscopic data. Since there are no known standard candles at redshifts  higher than $\sim$1.5, it is impossible to have an ``external'' test of the validity of the relation in a cosmology-independent way. The only way to confirm or falsify the method is through an analysis of the possible redshift dependent physical and/or selection effects that may bias the distance estimate. In this sense, the complete UV (rest-frame) spectral analysis is a significant step forward: possible biases related to the use of the optical broad-band magnitudes (for example, due to dust reddening, or to the effect of strong lines moving in or out of the photometric bands depending on the redshift) are ruled out, and the flux measurements are directly done from the spectral fits. 

The second fundamental outcome of our work is the first systematic search of the best X-ray and UV proxies of the relation, within the spectral range covered by the available spectroscopic data. While the best X-ray and UV indicators turned out to be the monochromatic flux at 1~keV and the Mg\,\textsc{ii} flux, respectively, we found that the intrinsic dispersion of the relation with the other proxies is only slightly larger. In particular, the ``standard'' indicators, i.e. the monochromatic flux at 2500~{\AA} derived from the photometric data, and the monochromatic flux at 2~keV are almost as effective as the ``best'' ones.

This result suggests that none of the indicators used here is the primary driver of the relation. In order to prove this statement, we can consider our best UV indicator, i.e. the Mg\,\textsc{ii} flux, and the intrinsic dispersion of the X-ray to UV relation, $\delta_{\rm Mg\,\textsc{ii}}$, obtained with this indicator. If the Mg\,\textsc{ii} flux were the primary driver of the relation (i.e. the physical relation involves either the Mg\,\textsc{ii} flux or a tightly related quantity) we would expect that the dispersion $\delta$ of the X-ray to UV relation using another UV proxy would be related to $\delta_{\rm Mg\,\textsc{ii}}$ through the following relation: $\delta^2 = \delta^2_{\rm Mg\,\textsc{ii}}+\Delta^2$, where $\Delta$ is the dispersion of the relation between the Mg\,\textsc{ii} flux and the other UV indicator. However, this is not the case: for example, the dispersion of the relation between the Mg\,\textsc{ii} flux and the monochromatic flux at 2500~{\AA}  is $\Delta=0.15$ dex, which would imply $\delta\sim0.25$ dex, while the observed value is $\delta=0.19\pm0.01$ dex (see Fig. \ref{gamma_mg_1kev}). We conclude that both the UV indicators used here are proxies of a more fundamental one, probably related to the UV emission blueward of the Lyman limit.

A final result with some implications for the use of quasars as standard candles is the relation between the X-ray and UV fluxes and the FWHM of the Mg\,\textsc{ii} emission line. 
We tested the inclusion of this parameter in the X-ray to UV flux relation, using the photometric 2500 {\AA} flux as $f_{\rm UV}$. Just like in the spectroscopic case, we found that there is no correlation between the UV flux and the Mg\,\textsc{ii} FWHM. Therefore, we can incorporate an additional term in the relation between fluxes when the FWHM of the Mg\,\textsc{ii} emission line is available while fitting the ``traditional'' relation if not. When doing so, we obtain $\gamma=0.58\pm0.01$, $\delta = 0.16$ dex, and $\delta_T = 0.19$ dex. 

\begin{figure}
	\includegraphics[scale=0.41]{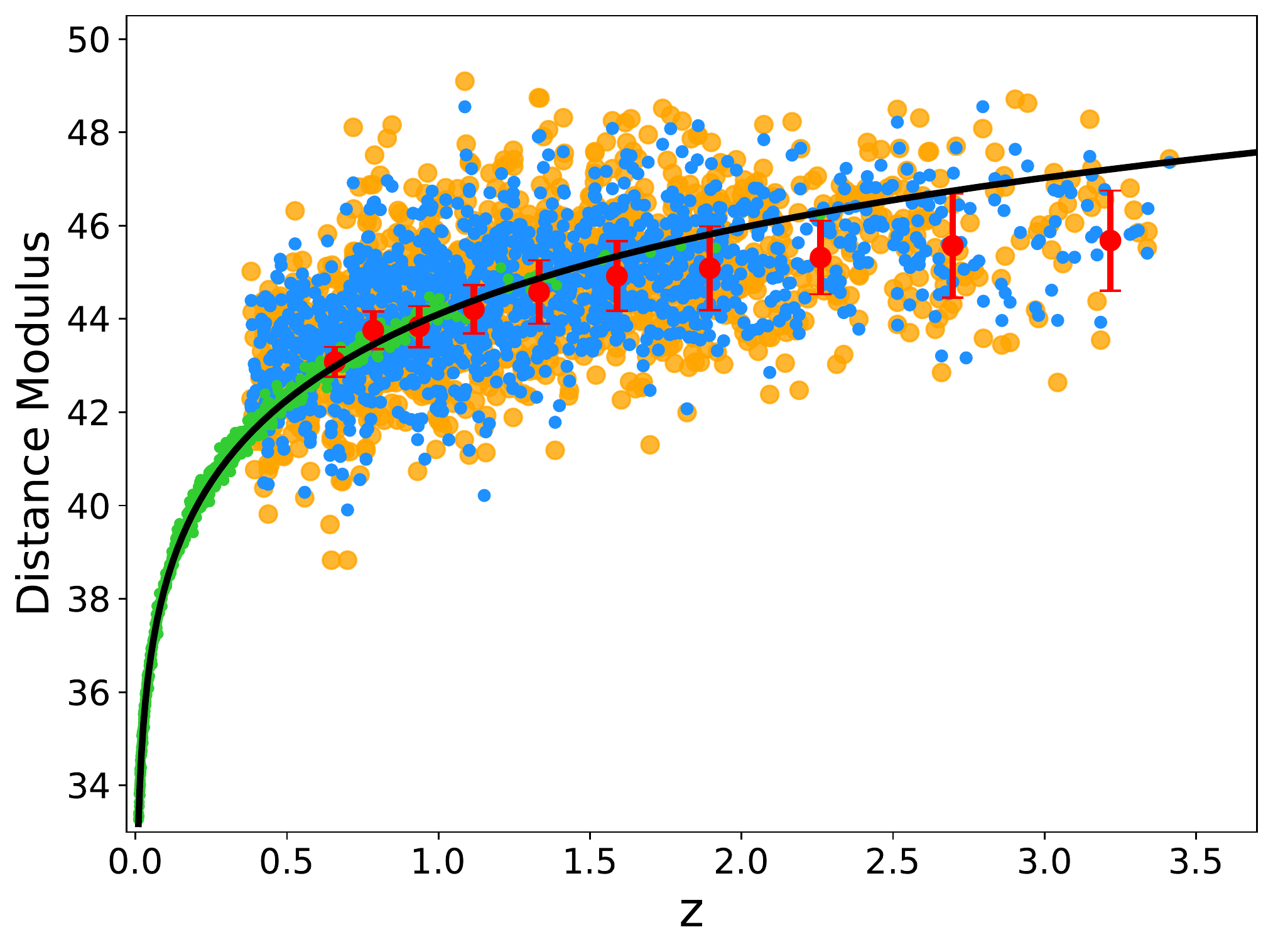}
	\centering
	\caption{Hubble diagram of supernovae and quasars: green points are supernovae Ia from the Pantheon sample \citep{Scolnic18}, yellow points are quasars with distances derived using the photometric UV fluxes, blue points are quasars with distances derived using the spectroscopic UV fluxes; red points are the average distance modulus values for ``spectroscopic'' quasars in the individual redshift bins. The normalisation parameter for quasars is chosen in order 
to match that of supernovae Ia. As in our previous studies \citep{RL19_nature}, we do so by cross-matching the Hubble diagram of quasars with that of supernovae in the common redshift range. The black line represents the prediction of a flat $\Lambda$CDM model with $\Omega_{\rm M} = 0.3$.
}
	\label{HD}
\end{figure}

We applied the results discussed above to the construction of a new Hubble diagram of quasars, based on the ``spectroscopic'' quasar sample described in this work.  We derived the luminosity distances using equation  \ref{leggefw} for objects for which the Mg\,\textsc{ii} line is available, and the standard relation for those for which it is not.  We used the 1-keV monochromatic flux as $f_{\rm X}$. As for $f_{\rm UV}$, we used both the ``spectroscopic'' and the ``photometric'' monochromatic flux at 2500 {\AA} (we also repeated the computation using the Mg\,\textsc{ii} flux, obtaining fully consistent results).  The results are shown in Fig.~\ref{HD}, where we also plot the supernovae of the Pantheon sample \citep{Scolnic18}, used for the absolute calibration.
\citet{Petrosian2022} argued that the analysis in redshift intervals may bias the cosmological analysis.
We stress that we never considered any binning (especially when considering redshifts) to perform the cosmological analysis in our previous works. The same consideration can be applied here as well. The distance modulus of each individual quasar is obtained from Equations 9 and 10, and the fits of the Hubble diagram are done marginalizing over the parameters $\gamma$ and $\beta$ of the relation. As a consequence, the red points shown in Figure~\ref{HD} are computed only after the best fit values of these parameters are obtained, and they are shown only for illustrative purposes.

The Hubble diagram in Fig.~\ref{HD} is relevant for two main reasons:\\
1) The two versions of the diagram, with the two different methods to derive the UV monochromatic flux, are in full agreement. This was not granted a priori: if some systematic effect related to the derivation of the photometric fluxes were present, it would have been revealed by the comparison with the spectroscopy-based values. The photometric fluxes are much easier to obtain, and indeed we always used these values in our previous works. Therefore, this result is also a validation of the Hubble diagrams in \citet{RL19_nature} and \citet{Lusso20}, which revealed a strong tension with the ``concordance'' flat $\Lambda$CDM model.\\
2) The dispersion in the Hubble diagram based on spectroscopic points is significantly lower than that based on photometric fluxes. This is the consequence of the different slopes of the X-ray to UV relation depending on which UV flux is used: $\gamma_{\rm phot}\sim0.6$ for the photometric fluxes and $\gamma_{\rm spec}\sim0.45$ for the spectroscopic ones. Considering the error propagation from the relation to the distance moduli plotted in Fig.~\ref{HD}, the main contribution to the error is the intrinsic dispersion of the relation, divided by a factor ($1-\gamma$). This term implies that the uncertainty on the distances derived from spectroscopic fluxes is lower that of distances based on photometric fluxes by a factor $(1- \gamma_{\rm phot})/(1- \gamma_{\rm spec})\sim0.75$.

The tension with the standard flat $\Lambda$CDM model of the spectroscopic sample discussed in this paper has a statistical significance of 3$\sigma$. This is lower than previously published results based on larger samples (\citealt{Lusso20}, \citealt{bargiacchi21}) but, again, it is the first result based on a sample with complete control on the derivation of the UV fluxes.

\section{Summary and conclusions}
In this paper, we present a thorough UV spectral analysis of 1764 quasars from the \citet{Lusso20} sample with the main aims to discuss the choice of the $\lx$ and $\luv$ indicators and to establish whether it is possible to further reduce the observed dispersion to gain a better understanding of the physics behind the $\lx-\luv$ relation. We thus derived monochromatic luminosities at five different wavelengths as well as the emission-line properties. We compared our results with the spectral analysis of \cite{Wu22} and we found a good agreement (see Appendix A). We also computed the X-ray properties of our sample from a spectroscopic analysis for objects at a redshift higher than 1.9, while we used photometric data for objects at lower redshift (see Appendix B).
Our main results are summarised below: 
\begin{enumerate}
    \item We obtained the monochromatic flux at 2500 {\AA} from a detailed spectroscopic analysis, as this is supposed to be a more accurate measurement than the one derived from the photometric data. We then analysed the $\lx-\luv$ relation in narrow redshift bins, so that we could (i) use fluxes instead of luminosities and therefore be independent of any chosen cosmology, and (ii) look for possible redshift trends of the slope parameter $\gamma$. While the slope is confirmed not to show any systematic trend with redshift, its value of $\gamma\sim 0.46$ is lower (flatter) than those found in our previous studies where photometric data were used. Also, the dispersion parameter $\delta$ is slightly higher when using spectroscopically-derived monochromatic fluxes as $f_{\rm UV}$. If the true physical quantity behind the $\lx-\luv$ relation had been the monochromatic luminosity at 2500  {\AA} (or any of the wavelengths that we tested, see below) we would have expected better results in terms of the dispersion. As this does not occur, it probably means that the spectroscopic and the photometric luminosities are simply two different \textit{proxies} of another quantity, and thus they are similarly effective when used in the X-ray to UV relation.
    \item We investigated what the best energies to be used as $\lx$ and $\luv$ are, following the assumption that both the UV and the X-ray continua of quasars can be parameterized as power laws and studying the dependence of the X-ray to UV relation on the respective spectral slopes. We stress that we do not expect the relation to subsist between two monochromatic luminosities, but we are looking for the best possible proxies for the overall UV (disc) and X-ray (corona) emission. In the X-rays, we find that there is a preference for 1 keV as the characteristic energy, i.e. the one less sensitive to the actual spectral slope. When using the 1-keV fluxes instead of the 2-keV ones, we found indeed a lower dispersion. Unfortunately, given the redshift range of our sample, the 1-keV flux is on average measured with higher uncertainties than the 2-keV ones. This partly undermines the advantage of having found the characteristic energy because the resulting total dispersion is only slightly lower. Still, the fact that the total dispersion decreases even if we are using a ``worse'' proxy in terms of uncertainties means that physically the relation is indeed tighter when we are using the 1-keV flux as $f_{\rm X}$. In the UV, we find no conclusive indications on a specific characteristic wavelength. This might be explained if the ``true'' physical quantity is found at much shorter wavelengths, beyond the peak of the disc emission, so that assuming the optical--UV power law is less appropriate and/or informative. Consequently, we argue that the best choice is the 2500 {\AA} flux simply because it is the one that results in a marginally lower dispersion. This might be because this wavelength falls in the observed spectra for the wider redshift range in our sample, implying a lower number of objects for which fluxes are determined via extrapolation and therefore with larger uncertainties.
    \item When using the integrated Mg\,\textsc{ii} line flux as $f_{\rm UV}$, we obtain a higher (steeper) slope value ($\gamma\sim0.6$) and a lower dispersion ($\delta=0.16$ dex) compared to any other UV continuum indicator derived from the spectroscopic analysis. We note that, although the Mg\,\textsc{ii} emission line is found at 2800 {\AA}, its flux strongly depends on the quasar emission at much shorter wavelengths, around $\sim$800~\AA. Therefore, it might be that the physical relation behind the X-ray and UV luminosities is more strongly linked to the quasar emission at shorter wavelengths and that, as a consequence, the Mg\,\textsc{ii} emission-line flux works as a better proxy when compared to the fluxes in the 1350--5500 {\AA} range.  Unfortunately, the Mg\,\textsc{ii} emission line is only available for the objects in our sample up to $z\sim2.5$. Another possible explanation is that, when we consider an indicator like the Mg\,\textsc{ii} emission line, which strongly depends upon the extreme-UV SED shape, a tighter correlation with the soft X-ray can naturally arise due to the energy proximity of the bands involved.
    \item The comparison between the values of the slopes that we found when using, respectively, the Mg\,\textsc{ii} line fluxes and the spectroscopic monochromatic fluxes, is entirely consistent with the presence of the Baldwin effect.
    \item  We confirm a correlation between the X-ray and UV flux taking into account the FWHM of the Mg\,\textsc{ii} line, whilst the UV flux and the FWHM turn out to be not significantly correlated. This non-correlation allows us to include the FWHM in the X-day/UV flux relation whenever available, and to keep using the standard relation otherwise. In this way, we can overcome the redshift limitations on the Mg\,\textsc{ii} flux and still obtain a lower dispersion for the whole sample.
    \item The Hubble diagram obtained from spectroscopic UV data is fully consistent with that obtained with photometric data. This is a validation of the previous results based on ``photometric'' Hubble diagrams. Moreover, the ``spectroscopic'' Hubble diagram shows a tension at stastistical level of $\sim$3$\sigma$ with the flat $\Lambda$CDM model. In previous works (e.g. \citealt{Lusso20}, \citealt{bargiacchi21}) we obtained a higher significance of this tension, thanks to a wider redshift extent than the sample considered here. However, the Hubble diagram presented here is the first one where we have fully checked the UV spectral properties of the sources (and also the X-ray ones at $z>1.9$).
\end{enumerate}

Overall, the results presented in this work are another step toward the validation of the non-linear X-ray to UV relation 
of quasars as a reliable distance indicator. Since a limited of number of supernovae Ia is available at redshifts higher than $\sim$1.5 and, by construction, a cosmology-independent validation of the method is impossible, the only way to further check our method is to search for possible evolutionary effects in the spectral emission of the quasars included in our Hubble diagram. We demonstrated in \cite{Sacchi22} that the average continuum and line properties of quasars at $z>2.5$ are perfectly matched to the ones of lower redshift counterparts in both the UV and X-rays. An extensive analysis of the stacked SDSS spectra in bins of BH mass and Eddington ratio for all the sources of the current sample is currently ongoing (Trefoloni et al.~2023, in preparation). We expect that future observations of supernovae at $z>1.5$ will be able to independently probe any deviation from the concordance model found with the Hubble diagram of quasars.


\begin{acknowledgements}
We thank the referee for their detailed and constructive suggestions. We acknowledge financial contribution from the agreement ASI-INAF n.2017-14-H.O. EL acknowledges the support of grant ID: 45780 Fondazione Cassa di Risparmio Firenze. 
\end{acknowledgements}

\bibliographystyle{aa} 
\bibliography{bibl}

\begin{thebibliography}{33}
\expandafter\ifx\csname natexlab\endcsname\relax\def\natexlab#1{#1}\fi

\bibitem[{{Arnaud}(1996)}]{Arnaud96}
{Arnaud}, K.~A. 1996, in Astronomical Society of the Pacific Conference Series,
  Vol. 101, Astronomical Data Analysis Software and Systems V, ed. G.~H.
  {Jacoby} \& J.~{Barnes}, 17

\bibitem[{{Baldwin}(1977)}]{Baldwin77}
{Baldwin}, J.~A. 1977, \apj, 214, 679

\bibitem[{{Bargiacchi} {et~al.}(2022){Bargiacchi}, {Benetti}, {Capozziello},
  {Lusso}, {Risaliti}, \& {Signorini}}]{bargiacchi22}
{Bargiacchi}, G., {Benetti}, M., {Capozziello}, S., {et~al.} 2022, \mnras, 515,
  1795

\bibitem[{{Bargiacchi} {et~al.}(2021){Bargiacchi}, {Risaliti}, {Benetti},
  {Capozziello}, {Lusso}, {Saccardi}, \& {Signorini}}]{bargiacchi21}
{Bargiacchi}, G., {Risaliti}, G., {Benetti}, M., {et~al.} 2021, \aap, 649, A65

\bibitem[{{Calderone} {et~al.}(2017){Calderone}, {Nicastro}, {Ghisellini},
  {Dotti}, {Sbarrato}, {Shankar}, \& {Colpi}}]{Calderone17}
{Calderone}, G., {Nicastro}, L., {Ghisellini}, G., {et~al.} 2017, \mnras, 472,
  4051

\bibitem[{{Elvis} {et~al.}(2012){Elvis}, {Hao}, {Civano}, {Brusa}, {Salvato},
  {Bongiorno}, {Capak}, {Zamorani}, {Comastri}, {Jahnke}, {Lusso}, {Mainieri},
  {Trump}, {Ho}, {Aussel}, {Cappelluti}, {Cisternas}, {Frayer}, {Gilli},
  {Hasinger}, {Huchra}, {Impey}, {Koekemoer}, {Lanzuisi}, {Le Floc'h}, {Lilly},
  {Liu}, {McCarthy}, {McCracken}, {Merloni}, {Roeser}, {Sanders}, {Sargent},
  {Scoville}, {Schinnerer}, {Schiminovich}, {Silverman}, {Taniguchi},
  {Vignali}, {Urry}, {Zamojski}, \& {Zatloukal}}]{Elvis2012}
{Elvis}, M., {Hao}, H., {Civano}, F., {et~al.} 2012, \apj, 759, 6

\bibitem[{{Foreman-Mackey} {et~al.}(2013){Foreman-Mackey}, {Hogg}, {Lang}, \&
  {Goodman}}]{emcee13}
{Foreman-Mackey}, D., {Hogg}, D.~W., {Lang}, D., \& {Goodman}, J. 2013, \pasp,
  125, 306

\bibitem[{{Haardt} \& {Maraschi}(1991)}]{Haardt91}
{Haardt}, F. \& {Maraschi}, L. 1991, \apjl, 380, L51

\bibitem[{{Haardt} \& {Maraschi}(1993)}]{Haardt93}
{Haardt}, F. \& {Maraschi}, L. 1993, \apj, 413, 507

\bibitem[{{Khadka} \& {Ratra}(2021)}]{khadka2021}
{Khadka}, N. \& {Ratra}, B. 2021, \mnras, 502, 6140

\bibitem[{{Krolik} \& {Kallman}(1988)}]{Krolik88}
{Krolik}, J.~H. \& {Kallman}, T.~R. 1988, \apj, 324, 714

\bibitem[{{Lusso} {et~al.}(2010){Lusso}, {Comastri}, {Vignali}, {Zamorani},
  {Brusa}, {Gilli}, {Iwasawa}, {Salvato}, {Civano}, {Elvis}, {Merloni},
  {Bongiorno}, {Trump}, {Koekemoer}, {Schinnerer}, {Le Floc'h}, {Cappelluti},
  {Jahnke}, {Sargent}, {Silverman}, {Mainieri}, {Fiore}, {Bolzonella}, {Le
  F{\`e}vre}, {Garilli}, {Iovino}, {Kneib}, {Lamareille}, {Lilly}, {Mignoli},
  {Scodeggio}, \& {Vergani}}]{Lusso10}
{Lusso}, E., {Comastri}, A., {Vignali}, C., {et~al.} 2010, \aap, 512, A34

\bibitem[{{Lusso} {et~al.}(2021){Lusso}, {Nardini}, {Bisogni}, {Risaliti},
  {Gilli}, {Richards}, {Salvestrini}, {Vignali}, {Bargiacchi}, {Civano},
  {Elvis}, {Fabbiano}, {Marconi}, {Sacchi}, \& {Signorini}}]{lussoCIV}
{Lusso}, E., {Nardini}, E., {Bisogni}, S., {et~al.} 2021, \aap, 653, A158

\bibitem[{{Lusso} \& {Risaliti}(2016)}]{RL16_tight}
{Lusso}, E. \& {Risaliti}, G. 2016, \apj, 819, 154

\bibitem[{{Lusso} \& {Risaliti}(2017)}]{LR17}
{Lusso}, E. \& {Risaliti}, G. 2017, \aap, 602, A79

\bibitem[{{Lusso} {et~al.}(2020){Lusso}, {Risaliti}, {Nardini}, {Bargiacchi},
  {Benetti}, {Bisogni}, {Capozziello}, {Civano}, {Eggleston}, {Elvis},
  {Fabbiano}, {Gilli}, {Marconi}, {Paolillo}, {Piedipalumbo}, {Salvestrini},
  {Signorini}, \& {Vignali}}]{Lusso20}
{Lusso}, E., {Risaliti}, G., {Nardini}, E., {et~al.} 2020, \aap, 642, A150

\bibitem[{{Meyer} {et~al.}(2000){Meyer}, {Liu}, \&
  {Meyer-Hofmeister}}]{Meyer2000}
{Meyer}, F., {Liu}, B.~F., \& {Meyer-Hofmeister}, E. 2000, \aap, 361, 175

\bibitem[{{Netzer}(1980)}]{Netzer80}
{Netzer}, H. 1980, \apj, 236, 406

\bibitem[{{Petrosian} {et~al.}(2022){Petrosian}, {Singal}, \&
  {Mutchnick}}]{Petrosian2022}
{Petrosian}, V., {Singal}, J., \& {Mutchnick}, S. 2022, \apjl, 935, L19

\bibitem[{{Rakshit} {et~al.}(2020){Rakshit}, {Stalin}, \&
  {Kotilainen}}]{griglia}
{Rakshit}, S., {Stalin}, C.~S., \& {Kotilainen}, J. 2020, \apjs, 249, 17

\bibitem[{{Richards} {et~al.}(2006){Richards}, {Lacy}, {Storrie-Lombardi},
  {Hall}, {Gallagher}, {Hines}, {Fan}, {Papovich}, {Vanden Berk}, {Trammell},
  {Schneider}, {Vestergaard}, {York}, {Jester}, {Anderson}, {Budav{\'a}ri}, \&
  {Szalay}}]{Richards06}
{Richards}, G.~T., {Lacy}, M., {Storrie-Lombardi}, L.~J., {et~al.} 2006, \apjs,
  166, 470

\bibitem[{{Risaliti} \& {Lusso}(2019)}]{RL19_nature}
{Risaliti}, G. \& {Lusso}, E. 2019, Nature Astronomy, 3, 272

\bibitem[{{Sacchi} {et~al.}(2022){Sacchi}, {Risaliti}, {Signorini}, {Lusso},
  {Nardini}, {Bargiacchi}, {Bisogni}, {Civano}, {Elvis}, {Fabbiano}, {Gilli},
  {Trefoloni}, \& {Vignali}}]{Sacchi22}
{Sacchi}, A., {Risaliti}, G., {Signorini}, M., {et~al.} 2022, \aap, 663, L7

\bibitem[{{Sanders} {et~al.}(1989){Sanders}, {Phinney}, {Neugebauer}, {Soifer},
  \& {Matthews}}]{Sanders89}
{Sanders}, D.~B., {Phinney}, E.~S., {Neugebauer}, G., {Soifer}, B.~T., \&
  {Matthews}, K. 1989, \apj, 347, 29

\bibitem[{{Scolnic} {et~al.}(2022){Scolnic}, {Brout}, {Carr}, {Riess}, {Davis},
  {Dwomoh}, {Jones}, {Ali}, {Charvu}, {Chen}, {Peterson}, {Popovic}, {Rose},
  {Wood}, {Brown}, {Chambers}, {Coulter}, {Dettman}, {Dimitriadis},
  {Filippenko}, {Foley}, {Jha}, {Kilpatrick}, {Kirshner}, {Pan}, {Rest},
  {Rojas-Bravo}, {Siebert}, {Stahl}, \& {Zheng}}]{Scolnic22}
{Scolnic}, D., {Brout}, D., {Carr}, A., {et~al.} 2022, \apj, 938, 113

\bibitem[{{Scolnic} {et~al.}(2018){Scolnic}, {Jones}, {Rest}, {Pan},
  {Chornock}, {Foley}, {Huber}, {Kessler}, {Narayan}, {Riess}, {Rodney},
  {Berger}, {Brout}, {Challis}, {Drout}, {Finkbeiner}, {Lunnan}, {Kirshner},
  {Sanders}, {Schlafly}, {Smartt}, {Stubbs}, {Tonry}, {Wood-Vasey}, {Foley},
  {Hand}, {Johnson}, {Burgett}, {Chambers}, {Draper}, {Hodapp}, {Kaiser},
  {Kudritzki}, {Magnier}, {Metcalfe}, {Bresolin}, {Gall}, {Kotak}, {McCrum}, \&
  {Smith}}]{Scolnic18}
{Scolnic}, D.~M., {Jones}, D.~O., {Rest}, A., {et~al.} 2018, \apj, 859, 101

\bibitem[{{Shakura} \& {Sunyaev}(1973)}]{SS73}
{Shakura}, N.~I. \& {Sunyaev}, R.~A. 1973, \aap, 24, 337

\bibitem[{{Steffen} {et~al.}(2006){Steffen}, {Strateva}, {Brandt}, {Alexander},
  {Koekemoer}, {Lehmer}, {Schneider}, \& {Vignali}}]{Steffen06}
{Steffen}, A.~T., {Strateva}, I., {Brandt}, W.~N., {et~al.} 2006, \aj, 131,
  2826

\bibitem[{{Svensson}(1982)}]{Svensson82}
{Svensson}, R. 1982, \apj, 258, 321

\bibitem[{{Svensson} \& {Zdziarski}(1994)}]{Svensson1994}
{Svensson}, R. \& {Zdziarski}, A.~A. 1994, \apj, 436, 599

\bibitem[{{Tananbaum} {et~al.}(1979){Tananbaum}, {Avni}, {Branduardi}, {Elvis},
  {Fabbiano}, {Feigelson}, {Giacconi}, {Henry}, {Pye}, {Soltan}, \&
  {Zamorani}}]{Tananbaum79}
{Tananbaum}, H., {Avni}, Y., {Branduardi}, G., {et~al.} 1979, \apjl, 234, L9

\bibitem[{{Wu} \& {Shen}(2022)}]{Wu22}
{Wu}, Q. \& {Shen}, Y. 2022, arXiv e-prints, arXiv:2209.03987

\bibitem[{{Young} {et~al.}(2010){Young}, {Elvis}, \& {Risaliti}}]{Young2010}
{Young}, M., {Elvis}, M., \& {Risaliti}, G. 2010, \apj, 708, 1388

\end{thebibliography}

\appendix
\section{UV Spectral analysis} 
We performed the fit of the SDSS UV spectra for the 1764 objects in our sample with the IDL package \texttt{QSFit} \cite{Calderone17}. This software allows us to fit the AGN continuum, the Balmer continuum, the emission-line and iron-complex properties, and the host-galaxy component. Regarding the host galaxy, a single template of an elliptical galaxy was used to determine its contribution to the total luminosity. The code is highly customizable, and our main settings are described below:
\begin{itemize}
	\item[--] For quasars with redshift below 0.6, given that the host-galaxy luminosity can be a relevant share of the total luminosity, the slope of the quasar continuum is degenerate with the host galaxy normalization. Therefore, there is no way to determine it in a reliable way. We consequently fixed the value of the continuum slope for such quasars as $f_\lambda \propto \lambda^{-1.7}$.
	\item[--] Each emission line was fitted with a Gaussian profile. \texttt{QSFit} allows us to consider both a broad and narrow component to fit each line, and to add unknown components if necessary. In case the code is not able to correctly disentangle the broad and the narrow component of the main emission lines, we used a composite profile of a broad component and an unknown one to account for the residuals. With this method, what we obtained are the \textit{total} emission line properties, rather than the separated properties of broad and narrow components. 
	\item[--] Full Width at Half Maximum (FWHM) upper limits were set to 2000 km/s for unknown and narrow lines and to 10,000 km/s for broad lines.
	\item[--] To avoid fitting regions of the spectrum contaminated by intergalactic absorption, the minimum wavelength for model fitting was set to 1450 {\AA} (rest frame) for all the objects in our sample.
\end{itemize}
The strength of the \texttt{QSFit} code is that the spectral properties are all fitted simultaneously. This way, one can reasonably assume that the obtained continuum properties do not dependent on local features of the spectrum itself. 

\begin{figure*}
	\includegraphics[scale=0.8]{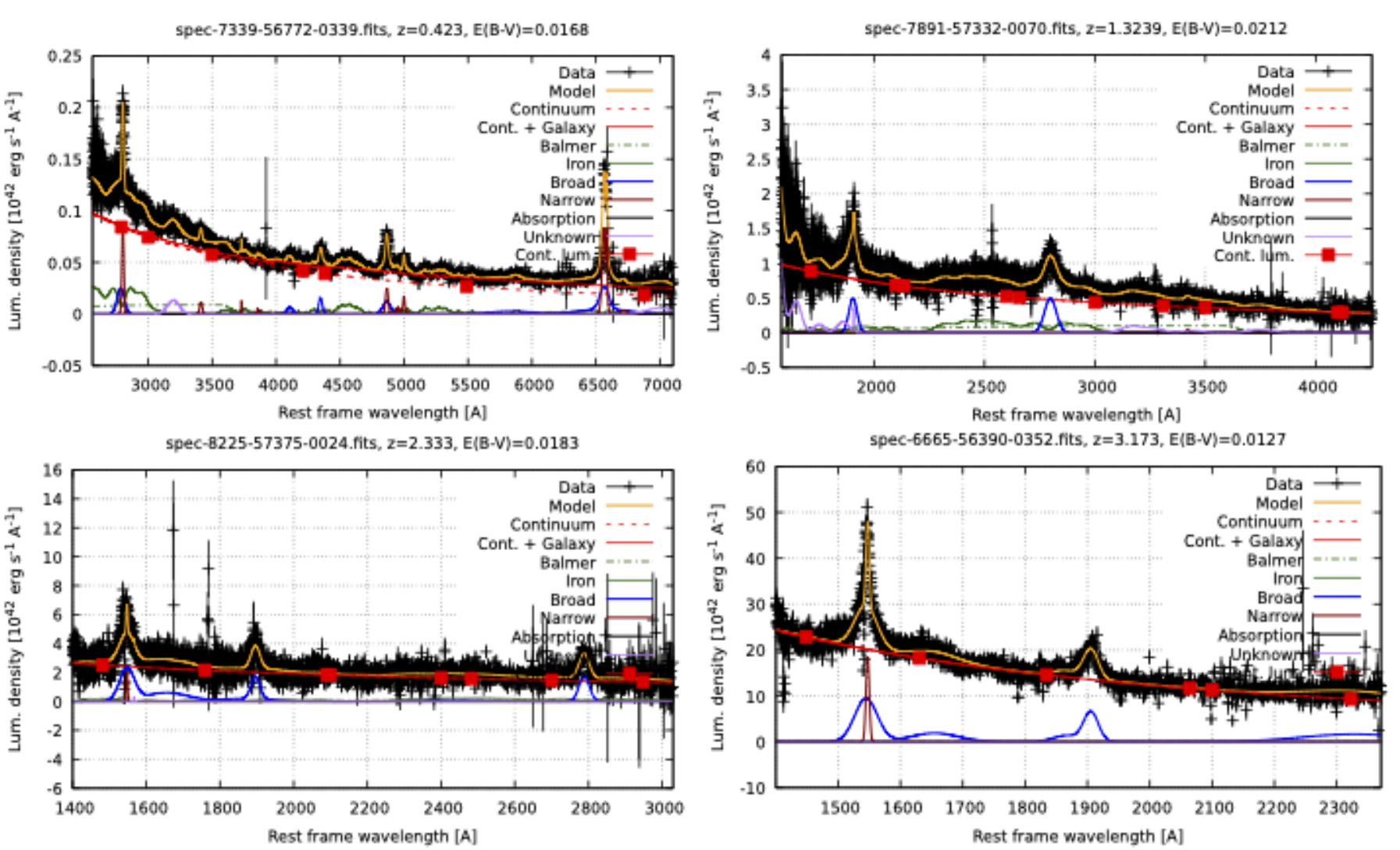}
	\centering
	\caption{Four spectra at different redshifts and \texttt{QSFit} best-fit results, in yellow. The different components are shown in the legend on the right of each panel.}
	\label{4horsemen}
\end{figure*}

At the end of the fitting procedure, for each quasar we obtained:
\begin{itemize}
   	\item[--] The continuum slope and the monochromatic luminosity at 2500 \AA. We also recorded the luminosity values at 1350 {\AA}, 3000 {\AA}, 4400 {\AA}, 5100 {\AA}.  If one or more of these wavelengths was out of the spectral range for a given spectrum, we extrapolated its value adopting the best-fit continuum slope. 
   	\item[--] Line properties: total flux, FWHM, offset velocity, equivalent width. Together with the line properties and their errors, quality flags for each line were given. 
   	\item[--] Quality flag: the \texttt{QSFit} code automatically raises a flag whenever one or more of the following situations occur: (i) the value of the continuum luminosity, a line's FWHM or its offset velocity is NaN or equal to zero; (ii) any of the previous quantities hits a boundary value in the fit; (iii) the relative uncertainty on the continuum luminosity is higher than 1.5; (iv) the relative uncertainty on the FWHM of a given line is higher than 2; (v) the uncertainty on the velocity offset is higher than 500 km s$^{-1}$. Objects with bad-quality flags were removed from the sample.
   \end{itemize}
After the fitting procedure, each spectrum was visually inspected and a second quality flag was raised if: (i) the residuals of the fitting procedure showed a systematic trend as a function of wavelength, (ii) the mean reduced $\chi^{2}$ value was higher than 2. Since in these cases the monochromatic luminosities estimates could not be considered reliable, such sources were removed from the sample. As a consequence, our sample size decreased from 1764 to 1705 sources.
Among them, 1217 also have good Mg\,\textsc{ii} emission line properties (line luminosity, FWHM, EW, velocity offset), while 403 have H$\beta$, 305 have [O\,\textsc{iii}]\,$\lambda$4959\AA, 291 have [O\,\textsc{iii}]\,$\lambda$5007\AA\ (202 have both the doublet components), 493 have C\,\textsc{iv}. 
In Figure \ref{4horsemen} we show, as an example, the spectra of four objects at different redshifts and the best fit results of the \texttt{QSFit} analysis.

\begin{figure}
	\includegraphics[scale=0.6]{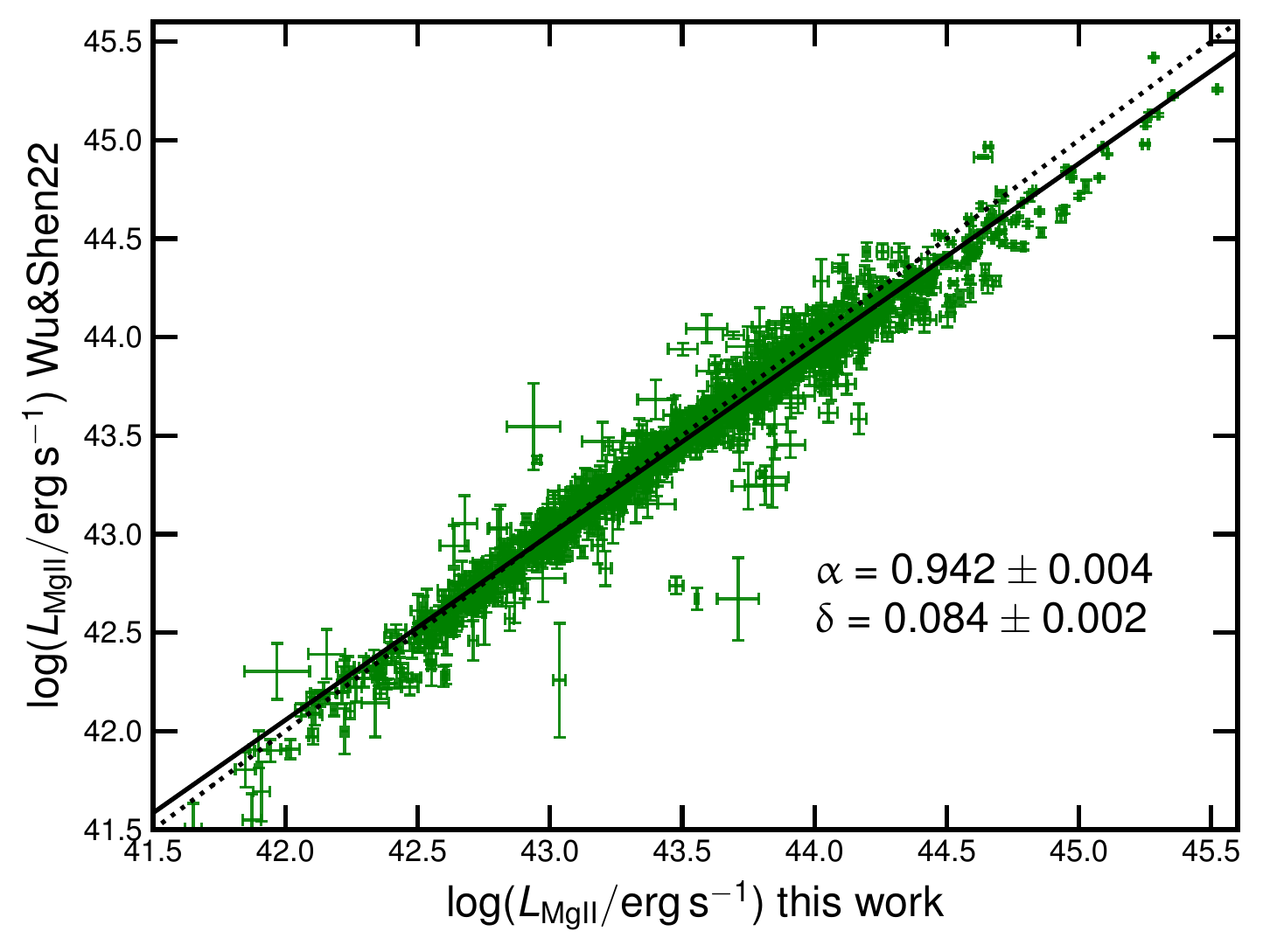}
	\centering
	\caption{Comparison between the Mg\,\textsc{ii} luminosity as derived from \cite{Wu22} and the Mg\,\textsc{ii} luminosity derived in this work, in logarithmic units (erg s$^{-1}$). The dotted line is the one-to-one relation. We also report the best fit slope and dispersion and the resulting best fit-regression line as the solid black line.}
	\label{grifw}
\end{figure}

In Figure \ref{grifw} we compare our estimates of the total luminosity of the Mg\,\textsc{ii} line with the one derived from the \cite{Wu22} catalogue. We can see an overall good agreement. Comparisons between other emission line properties show similar results. 
We also performed the same comparisons with the results of \cite{griglia}, and they gave analogous results. 

\begin{figure}
	\includegraphics[scale=0.6]{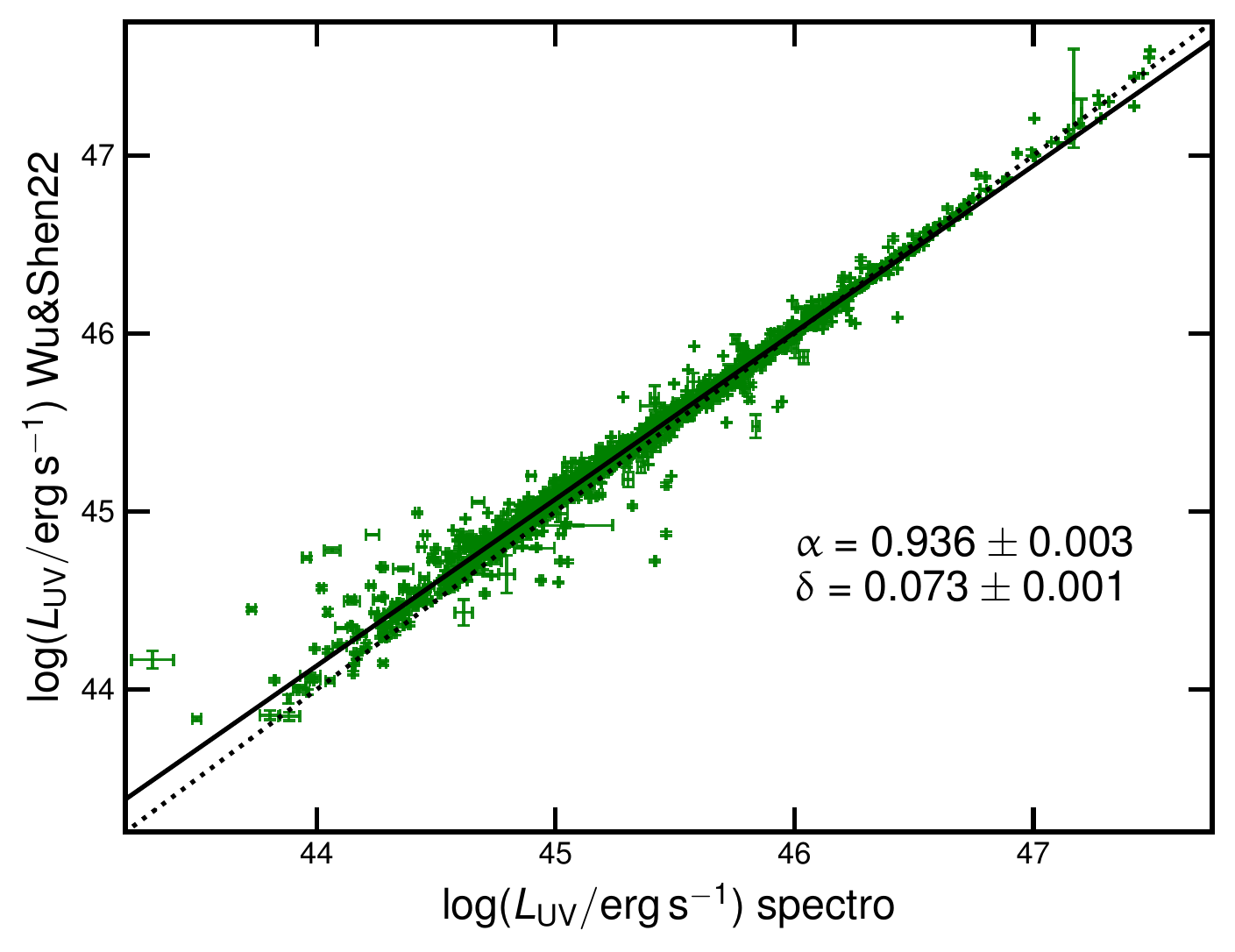}
	\centering
	\caption{Comparison between the monochromatic luminosity at 2500 {\AA} obtained from our spectral analysis and the one from the \cite{Wu22}, in logarithmic units (erg s$^{-1}$). The dotted line is the one-to-one relation. We also report the best -it slope and dispersion and the resulting best-fit regression line with the solid black line. There is an overall good agreement between the two measurements.}
	\label{g_3000}
\end{figure}
\label{fitprod}	

\cite{Wu22} presented the detailed measurements of the spectral properties of $\sim$500,000 quasars from the latest release of the Sloan Digital Sky Survey (SDSS-DR16) quasar catalogue, which have been used to validate the results of our spectral analysis. 
Regarding the monochromatic luminosities, we find a very good match, as can be seen in Figure \ref{g_3000} for the monochromatic luminosity at 3000 {\AA}. We can also see that the match is excellent at high luminosities, while some scatter is present for less luminous objects. This mainly depends on the different host-galaxy fitting techniques that have been employed: we used a single host-galaxy template, while in \cite{Wu22} the galaxy is not characterized. Quasars with lower luminosities are also the ones more affected by the host-galaxy contribution, so we expect higher discrepancies for such objects. Comparisons with the other monochromatic luminosities at different wavelengths (1350 {\AA}, 4400 {\AA}, 5100 {\AA}) all show analogous results. \\

\section{X-ray spectral analysis}
All the objects in our sample have an estimate of the rest-frame 2-keV monochromatic flux derived from {\it XMM-Newton} photometric data. A detailed description of the procedure can be found in Section 4 of \cite{Lusso20}. \\
We have performed a full spectroscopic analysis for a subsample of 292 objects, which are all the objects in our sample at redshift higher than 1.9, for which any discrepancy between the photometric and spectroscopic values can have major consequences in terms of cosmological applications.
This comes as an extension of the X-ray spectroscopic analysis provided in \cite{Sacchi22} for objects at redshift higher than 2.5. \\
The goal of this analysis is to prove that not using fully spectroscopic X-ray fluxes does not introduce any bias in our results, while showing at the same time  that part of the residual observed dispersion of the $\lx-\luv$ relation can be attributed to the lower accuracy of photometric measurements. We followed the standard procedure from the \textit{ XMM-Newton} user manual to obtain the spectra. For each object, we extracted three spectra for the three \textit{ XMM-Newton} cameras (pn, MOS1 and MOS2). We then combined the two MOS spectra into a single one. 

\begin{figure*}
	\includegraphics[scale=0.8]{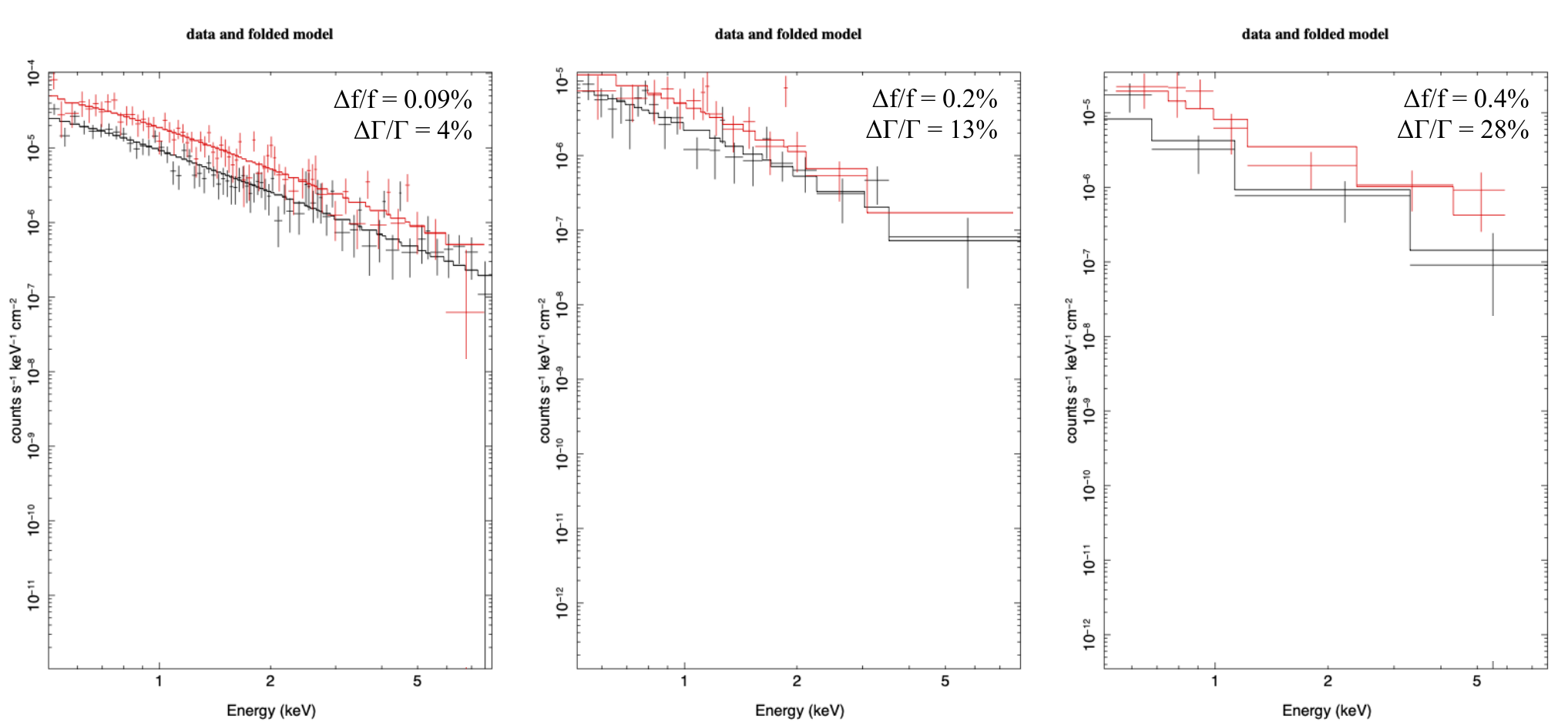}
	\centering
	\caption{Example of the X-ray spectrum and best fit of three sources at redshift 2.138, 2.144, 2.293. The pn spectrum is shown in red, and the MOS spectrum is shown in black. The relative uncertainties on the free parameters (the monochromatic flux at 2 keV and the photon index $\Gamma$) are shown as well.}
	\label{x_example}
\end{figure*}

\begin{figure}
	\includegraphics[scale=0.2]{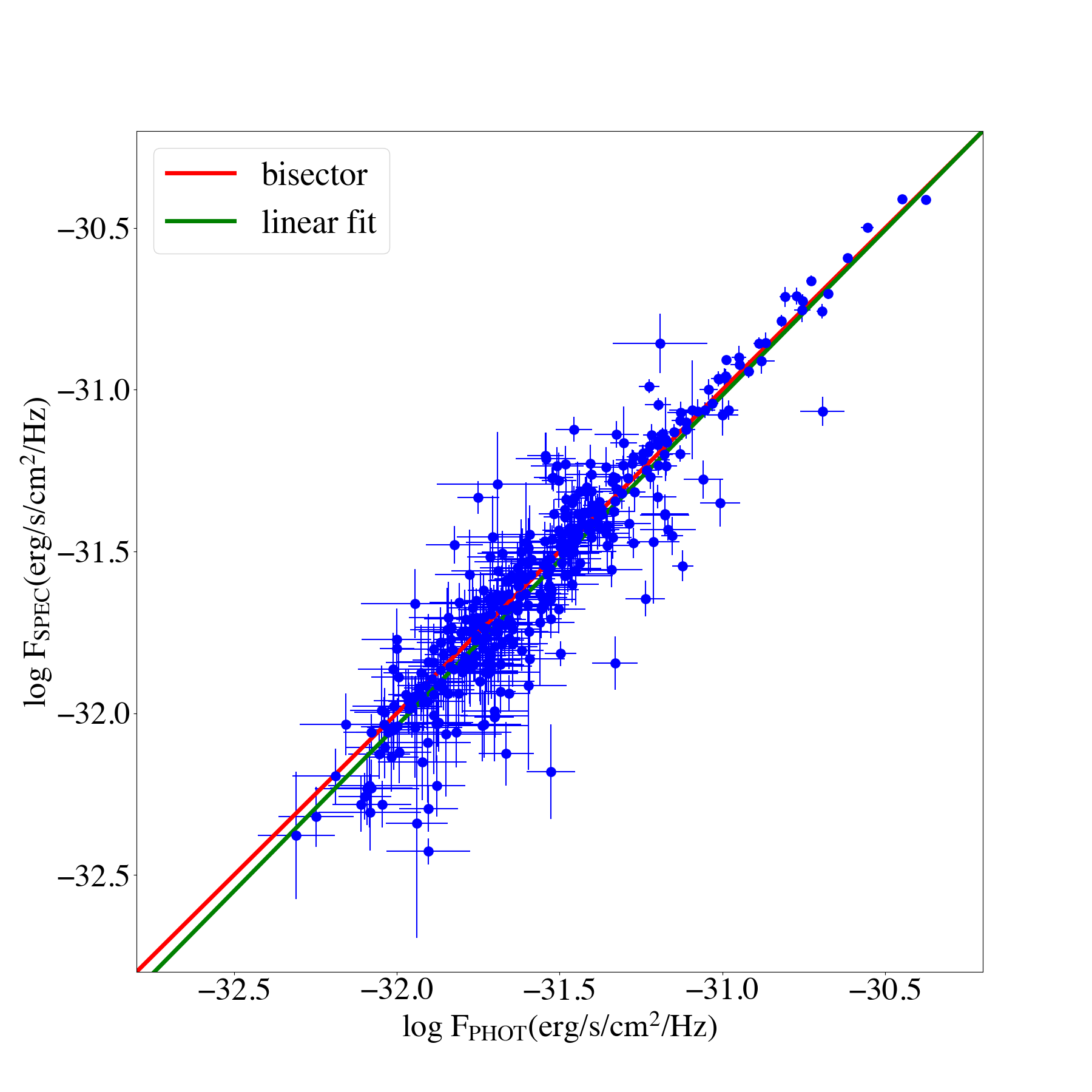}
	\centering
	\caption{Comparison between photometrically-derived 2-keV monochromatic fluxes and spectroscopically-derived ones. The red line represents the one-to-one relation, while the green line is the best fit of the relation between the two quantities.}
	\label{x_spec_phot}
\end{figure}

\begin{figure}
	\includegraphics[scale=0.6]{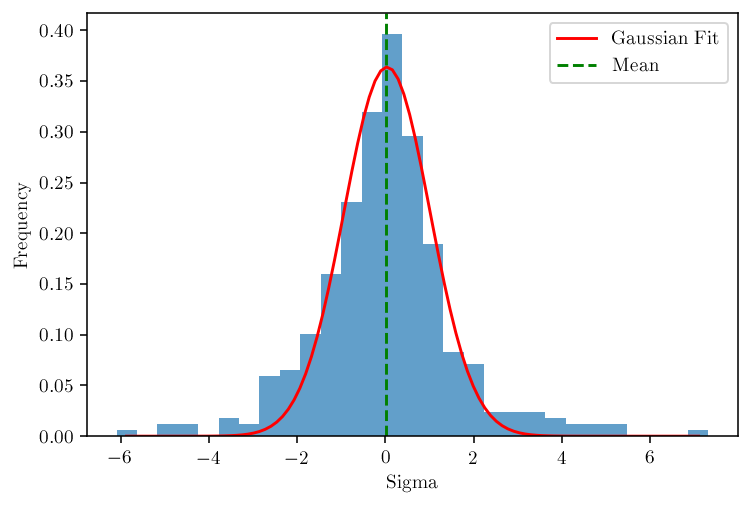}
	\centering
	\caption{Histogram of the differences between the photometric and the spectroscopic X-ray fluxes at 2 keV, shown in units of standard deviations. The red line shows the results of a Gaussian fit, which shows that the distribution is centered around zero. There is no significant skewness, as the skewness parameter turns out to be $k = 0.27\pm0.30$. This shows that there is no systematic shift between the two quantities.}
	\label{x_hist}
\end{figure}

The fit procedure was performed with the package XSPEC version 12.12 \citep{Arnaud96}. We subtracted the background from the spectrum and then fitted it with a power-law model, considering Galactic photoelectric absorption. We fitted the pn and the MOS spectra simultaneously, imposing the same spectral shape and allowing for a varying normalization constant between the two cameras. From the best fit we estimated the monochromatic flux at 1 keV, the monochromatic flux at 2 keV, and the photon index $\Gamma$, together with their 1$\sigma$ uncertainties. In Figure \ref{x_example} we show, as an example, the spectra and the best-fit model for three objects, with data of different quality and therefore different uncertainties on the determination of the monochromatic X-ray flux at 2 keV. 
In Figure \ref{x_spec_phot} we show the comparison between the monochromatic fluxes at 2 keV derived from this spectroscopic analysis with the ones obtain with the standard procedure. The relation between the two quantities was fitted with a linear relation and the best fit is statistically consistent with a one-to-one relation, with the linear regression returning a slope of $m=1.01\pm0.01$ as the best fit. In Figure \ref{x_hist} we also show the histogram of the differences between the spectroscopic and the monochromatic 2 keV fluxes, expressed in units of the standard deviation $\sigma$. We see also from this distribution that there is no systematic shift between the two quantities. Neither there is a significant skewness of the distribution, the skewness parameter being $k = 0.27\pm0.30$.

From this comparison, we infer that by using the spectroscopic data only for a sub-sample of objects we are not introducing any systematics. At the same time, there is a significant scatter between the spectroscopic and the photometric flux estimates. We can assume that the spectroscopic analysis is the more accurate one. This means that for those objects for which we are (still) using photometric data, we are actually introducing a contribution to the total observed dispersion when fitting the $\lx-\luv$ relation. We should also note that from the complete X-ray spectroscopic analysis we are also able to check one by one that the observations are not affected by any bias, which is a possible cause of additional dispersion. A discussion of the contribution of X-ray observational issues to the total observed dispersion of the $\lx-\luv$ relation will be addressed in a forthcoming paper. 
Analogous results in terms of the comparison between photometric and spectroscopic data are found when comparing the 1-keV monochromatic fluxes. 

\end{document}